\documentclass[10pt]{mpi08}
\pdfoutput=1
\usepackage{graphicx}
\usepackage{cite,./mcite} 
\usepackage{amsmath}  
\usepackage{relsize}  
\usepackage{booktabs} 

\def\lep{LEP}
\def\lepi{LEP~1}
\def\sld{SLD}
\def\petra{PETRA}
\def\rhic{RHIC}
\def\lhc{LHC}
\def\tevatron{Tevatron}
\def\runi{Run-\textsmaller{I}}
\def\runii{Run-\textsmaller{II}}

\def\delphi{Delphi}
\def\opal{Opal}
\def\cdf{CDF}
\def\dzero{D\O}

\def\professor{Professor}
\def\rivet{Rivet}
\def\pythia{Pythia}
\def\pythiasix{Pythia~6.4}

\def\PARJ#1{\ensuremath{\text{PARJ(#1)}}}
\def\PARP#1{\ensuremath{\text{PARP(#1)}}}
\def\MSTJ#1{\ensuremath{\text{MSTJ(#1)}}}
\def\MSTP#1{\ensuremath{\text{MSTP(#1)}}}

\def\gev{Ge\kern-0.035em V}
\def\tev{Te\kern-0.035em V}
\def\Nn#1#2{\ensuremath{N_{\,#1}^{\mspace{0.5mu}(\mspace{-0.5mu}#2)}}}
\def\Nmin#1{\ensuremath{N_\text{min}^{\mspace{0.5mu}(\mspace{-0.5mu}#1)}}}
\def\gof{GoF}
\def\p{\ensuremath{\vec{p}}}
\def\pT{\ensuremath{p_\perp}}

\def\Tab#1{Tab.~\ref{#1}}
\def\Tabs#1{Tables~\ref{#1}}
\def\Fig#1{Fig.~\ref{#1}}
\def\Figs#1{Figures~\ref{#1}}

\begin{document}

\let\unprotectedcite\cite
\def\cite{\protect\unprotectedcite}

\title{Monte Carlo tuning and generator validation}
\author{Andy Buckley$^1$,
        Hendrik Hoeth$^2$\protect\footnote{\ \ speaker} ,
        Heiko Lacker$^3$,
        Holger Schulz$^3$,
        Eike von Seggern$^3$}

\institute{%
$^1$Institute for Particle Physics Phenomenology, Durham University, UK\\
$^2$Department of Theoretical Physics, Lund University, Sweden\\
$^3$Physics Department, Berlin Humboldt University, Germany}
\maketitle

\begin{abstract}
  We present the Monte Carlo generator tuning strategy followed, and the
  tools developed, by the MCnet CEDAR project. We also present new
  tuning results for the \pythiasix{} event generator which are based on
  event shape and hadronisation observables from $e^+e^-$ experiments,
  and on underlying event and minimum bias data from the \tevatron{}.
  Our new tunes are compared to existing tunes and to Peter Skands' new
  ``Perugia'' tunes.
\end{abstract}

\section{Introduction}

With the \lhc{} starting soon, collider based particle physics is about
to enter a new energy regime. Everybody is excited about the
possibilities of finding new physics beyond the \tev{} scale, but the
vast majority of events at the \lhc{} will be Standard Model QCD events.
The proton will be probed at low Bj\"orken $x$ where current PDF fits
have large uncertainties, jets above 1\,\tev{} will be seen, and the
behaviour of the $pp$ total cross-section and multiple parton
interactions will be measured at values of $\sqrt{s}$ where
extrapolation from current data is challenging. No discoveries of new
physics can be claimed before the Standard Model at these energies is
measured and understood.

Monte Carlo event generators play an important role in virtually every
physics analysis at collider experiments. They are used to evaluate
signal and background events, and to design the analyses. It is
essential that the simulations describe the data as accurately as
possible. The main point here is not to focus on just one or two
distributions, but to look at a wide spectrum of observables. Only if
the Monte Carlo agrees with many complementary observables can we trust
it to have predictive power, and from disagreements we can learn
something about model deficiencies and the underlying physics.

As Monte Carlo event generators are based on phenomenological models and
approximations, there are a number of parameters that need to be tweaked
if the generator is to describe the experimental data. In the first part
of this talk we present a strategy for systematic Monte Carlo parameter
tuning. In the second part two new tunes of the \pythiasix{}
generator~\cite{Sjostrand:2006za} are presented and compared to other
tunings.

\section{MC tuning}

Every Monte Carlo event generator has a number of relatively free
parameters which must be tuned to make the generator describe
experimental data in the best possible way. Such parameters can be found
almost everywhere in Monte Carlo generators~-- all the way from the
(perturbative) hard interaction to the (non-perturbative) hadronisation
process. Naturally the majority of parameters are found in the
non-perturbative physics models.

While all the parameters have a physical motivation in their models,
there are usually only rough arguments about their scale. Other
parameters are measured experimentally (like $\alpha_s$), but as the
Monte Carlo event generators use them in a fixed-order scheme (unlike
nature) they need to be adjusted, too.

Going through the steps of event generation and identifying the most
important parameters, one typically finds $\mathcal{O}(20\mbox{--}30)$
parameters of particular importance to collider experiments. Most of
these parameters are highly correlated in a non-trivial way. We can
group the parameters in approximately independent sets e.\,g. in
flavour, fragmentation, and underlying event parameters, to reduce the
number to be optimised against any single set of observables.
Nevertheless, the number of parameters to be simultaneously tuned is
$\mathcal{O}(10)$. A manual or brute-force approach to Monte Carlo
tuning is not very practical: it is very slow, and manual tunings in
particular depend very much on the experience of the person performing
the tuning (at the same time there is a strong anti-correlation between
experience and willingness to produce a new tune manually).

\subsection{A systematic tuning strategy}

In this talk, we describe the \professor{} tuning system, which
eliminates the problems with manual and brute-force tunings by
parameterising a generator's response to parameter shifts on a
bin-by-bin basis, a technique introduced by the
\delphi-collaboration\cite{Hamacher:1995df, Abreu:1996na}. This
parameterisation, unlike a brute-force method, is then amenable to
numerical minimisation within a timescale short enough to make
explorations of tuning criteria possible.

\subsubsection{Predicting the Monte Carlo output}

The first step of any tuning is to define the parameters that shall be
varied, together with the variation intervals. This requires a thorough
understanding of the generator's model, its parameters and the available
data~-- all the relevant parameters for a certain model should enter the
tuning, but none of the irrelevant ones. A fragmentation tune for
example must include the shower cut-off parameter, while a tune of the
flavour composition had better not be dependent on it.

Once we have settled on a set of parameter intervals, it is time to
obtain a predictive function for the Monte Carlo output. Actually we
generate an ensemble of such functions. For each observable bin $b$ a
polynomial is fitted to the Monte Carlo response $\text{MC}_b$ to
changes in the parameter vector $\p=(p_1,\dots,p_P)$ of the $P$
parameters varied in the tune. To account for lowest-order parameter
correlations, a polynomial of at least second-order is used as the basis
for bin parameterisation:
\begin{align}
  \text{MC}_b(\p)
  \approx f^{(b)}(\p)
  = \alpha^{(b)}_0 + \sum_i \beta^{(b)}_i \, p^{}_i
  + \sum_{i \le j} \gamma^{(b)}_{ij} \, p^{}_i \, p^{}_j
\end{align}
We have tested this to give a good approximation of the true Monte Carlo
response for real-life observables.

The number of parameters and the order of the polynomial fix the number
of coefficients to be determined. For a second order polynomial in $P$
parameters, the number of coefficients is
\begin{align}
  \Nn{2}{P} = 1 + P + P(P+1)/2,
\end{align}
since only the independent components of the matrix term are to be
counted.

Given a general polynomial, we must now determine the coefficients
$\alpha,\beta,\gamma$ for each bin so as to best mimic the true
generator behaviour. This could be done by a Monte Carlo numerical
minimisation method, but there would be a danger of finding sub-optimal
local minima, and automatically determining convergence is a potential
source of problems. Fortunately, this problem can be cast in such a way
that a deterministic method can be applied.

One way to determine the polynomial coefficients would be to run the
generator at as many parameter points, $N$, as there are coefficients to
be determined. A square $N \times N$ matrix can then be constructed,
mapping the appropriate combinations of parameters on to the
coefficients to be determined; a normal matrix inversion can then be
used to solve the system of simultaneous equations and thus determine
the coefficients. Since there is no reason for the matrix to be
singular, this method will always give an ``exact'' fit of the
polynomial to the generator behaviour. However, this does not reflect
the true complexity of the generator response: we have engineered the
exact fit by restricting the number of samples on which our
interpolation is based, and it is safe to assume that taking a larger
number of samples would show deviations from what a polynomial can
describe, both because of intrinsic complexity in the true response
function and because of the statistical sampling error that comes from
running the generator for a finite number of events. What we would like
is to find a set of coefficients (for each bin) which average out these
effects and are a least-squares best fit to the oversampled generator
points. As it happens, there is a generalisation of matrix inversion to
non-square matrices --~the \emph{pseudoinverse} \cite{nla.cat-vn441566}
--~with exactly this property.

As suggested, the set of anchor points for each bin are determined by
randomly sampling the generator from $N$ parameter space points in the
$P$-dimensional parameter hypercube $[\,\p_{\text{min}},
\p_{\text{max}}]$ defined by the user. This definition requires physics
input~-- each parameter $p_i$ should have its upper and lower sampling
limits $p_{\text{min,max}}$ chosen so as to encompass all reasonable
values; we find that generosity in this definition is sensible, as
\professor{} may suggest tunes which lie outside conservatively chosen
ranges, forcing a repeat of the procedure. On the other hand the
parameter range should not be too large, in order to keep the volume of
the parameter space small and to make sure that the parabolic
approximation gives a good fit to the true Monte Carlo response. Each
sampled point may actually consist of many generator runs, which are
then merged into a single collection of simulation histograms. The
simultaneous equations solution described above is possible if the
number of sampled points is the same as the number of coefficients
between the $P$ parameters, i.e.  $N = \Nmin{P} = \Nn{n}{P}$. The more
robust pseudoinverse method applies when $N > \Nmin{P}$: we prefer to
oversample by at least a factor of 2. The numerical implementation of
the pseudoinverse uses a standard singular value decomposition (SVD)
\cite{1480176}.

\subsubsection{Comparing to data and optimising the parameters}

With the functions $f^{(b)}(\p)$ we now have a very fast way of
predicting the behaviour of the Monte Carlo generator. To get the Monte
Carlo response for any parameter setting inside the defined parameter
hypercube it is not necessary anymore to run the generator, but we can
simply evaluate the polynomial. This allows us to define a goodness of
fit function comparing data and (approximated) Monte Carlo which can be
minimised in a very short time.

We choose a heuristic $\chi^2$ function, but other goodness of fit
(\gof{}) measures can certainly be used. Since the relative importance
of various distributions in the observable set is a subjective thing
--~given 20 event shape distributions and one charged multiplicity, it
is certainly sensible to weight up the multiplicity by a factor of at
least 10 or so to maintain its relevance to the \gof{} measure~-- we
include per-observable weights, $w_{\mathcal{O}}$ for each observable
$\mathcal{O}$, in our $\chi^2$ definition:
\begin{align}
  \chi^2(\p) =
  \sum_{\mathcal{O}} w_{\mathcal{O}} \sum_{b \, \in \, \mathcal{O}}
  \frac{ (f^{(b)}(\p) - \mathcal{R}_b)^2 }{ \Delta^2_b },
\end{align}
where $\mathcal{R}_b$ is the reference (i.\,e. data) value for bin $b$
and the total error $\Delta_b$ is the sum in quadrature of the reference
error and the statistical generator errors for bin $b$. In practice we
attempt to generate sufficient events at each sampled parameter point
that the statistical MC error is much smaller than the reference error
for all bins.

It should be noted that there is unavoidable subjectivity in the choice
of these weights, and a choice of equal weights is no more sensible than
a choice of uniform priors in a Bayesian analysis; physicist input is
necessary in both choosing the admixture of observable weights according
to the criteria of the generator audience --~a $b$-physics experiment
may prioritise distributions that a general-purpose detector
collaboration would have little interest in~-- and to ensure that the
end result is not overly sensitive to the choice of weights.

The final stage is to minimise the parameterised
$\chi^2$ function. It is tempting to think that there is scope for an
analytic global minimisation at this order of polynomial, but not enough
Hessian matrix elements may be calculated to constrain all the
parameters and hence we must finally resort to a numerical minimisation.
This is the numerically weakest point in the method, as the weighted
quadratic sum of hundreds of polynomials is a very complex function and
there is scope for getting stuck in a non-global minimum. Hence the
choice of minimiser is important.

The output from the minimisation is a vector of parameter values which,
if the parameterisation and minimisation stages are faithful, should be
the optimal tune according to the (subjective) criterion defined by the
choice of observable weights.

\subsection{Tools}

We have implemented the tuning strategy described above in the
\professor{} software package. \professor{} reads in Monte Carlo and
data histogram files, parameterises the Monte Carlo response, and
performs the $\chi^2$ minimisation.

The Monte Carlo histograms used as input for \professor{} are generated
with \rivet{}~\cite{Buckley:2009ad}. \rivet{} is an analysis framework
for Monte Carlo event generator validation. By reading in HepMC event
records, \rivet{} can be used with virtually all common event
generators, and this well-defined interface between generator and
analysis tool ensures that the physics analyses are implemented in a
generator-independent way. A key feature of \rivet{} is that the
reference data can be taken directly from the HepData archive
\cite{Buckley:2006np} and is used to define the binnings of the Monte
Carlo histograms, automatically ensuring that there is no problem with
synchronising bin edge positions. At present, there are about 40 key
analyses mainly from \lep{} and \tevatron{}, but also from \sld{},
\rhic{}, \petra{}, and other accelerators. More analyses are constantly
being added.

\section{Tuning \pythiasix{}}

For the first production tuning we chose the \pythiasix{} event
generator, as this is a well-known generator which has been tuned before
and which we expected to behave well. Naturally the first step in tuning
a generator is to fix the flavour composition and the fragmentation
parameters to the precision data from \lep{} and \sld{} before
continuing with the parameters related to hadron collisions, for which we
use data from the \tevatron{}.

\subsection{Parameter factorisation strategy}

In \pythia{} the parameters for flavour composition decouple well from
the non-flavour hadronisation parameters such as $a$, $b$, $\sigma_q$, or
the shower parameters ($\alpha_s$, cut-off). Parameters related to the
underlying event and multiple parton interactions are decoupled from the
flavour and fragmentation parameters. In order to keep the number of
simultaneously tuned parameters small, we decided to follow a
three-stage strategy. In the first step the flavour parameters were
optimised, keeping almost everything else at its default values
(including using the virtuality-ordered shower). In the second step the
non-flavour hadronisation and shower parameters were tuned~-- using the
optimised flavour parameters obtained in the first step. The final step
was tuning the underlying event and multiple parton interaction
parameters to data from \cdf{} and \dzero{}.

\subsection{Flavour parameter optimisation}

\begin{table}
\begin{center}
\begin{tabular}{lrrl}
\toprule
Parameter  & Pythia 6.418 default & Final tune  &            \\
\midrule
\PARJ{1}   & 0.1       & 0.073 & diquark suppression         \\
\PARJ{2}   & 0.3       & 0.2   & strange suppression         \\
\PARJ{3}   & 0.4       & 0.94  & strange diquark suppression \\
\PARJ{4}   & 0.05      & 0.032 & spin-1 diquark suppression  \\
\PARJ{11}  & 0.5       & 0.31  & spin-1 light meson          \\
\PARJ{12}  & 0.6       & 0.4   & spin-1 strange meson        \\
\PARJ{13}  & 0.75      & 0.54  & spin-1 heavy meson          \\
\PARJ{25}  & 1         & 0.63  & $\eta$ suppression          \\
\PARJ{26}  & 0.4       & 0.12  & $\eta'$ suppression         \\
\bottomrule
\end{tabular}
\end{center}
\caption{Tuned flavour parameters and their defaults.}
\label{tab:tune-flavour}
\end{table}

The observables used in the flavour tune were hadron multiplicities and
their ratios with respect to the $\pi^+$ multiplicity measured at
\lepi{} and \sld{}~\cite{Amsler:2008zz}, as well as the $b$-quark
fragmentation function measured by the \delphi{}
collaboration~\cite{delphi-2002}, and flavour-specific mean charged
multiplicities as measured by the \opal{}
collaboration~\cite{Ackerstaff:1998hz}. For this first production we
chose to use a separate tuning of the Lund-Bowler fragmentation function
for $b$-quarks (invoked in \pythiasix{} by setting
$\MSTJ{11}=5$) with a fixed value of $r_b = 0.8$
(\PARJ{47}), as first tests during the validation phase of the
\professor{} framework showed that this setting yields a better
agreement with data than the default common Lund-Bowler parameters
for $c$ and $b$ quarks.

For the tuning we generated 500k~events at each of 180~parameter points.
The tuned parameters are the basic flavour parameters like diquark
suppression, strange suppression, or spin-1 meson rates. All parameters
are listed in \Tab{tab:tune-flavour} together with the tuning results.

Since the virtuality-ordered shower was used for tuning the flavour
parameters, we tested our results also with the \pT{}-ordered shower in
order to check if a separate tuning was necessary. Turning on the
\pT{}-ordered shower and setting $\Lambda_\text{QCD} = 0.23$ (the
recommended setting before our tuning effort) we obtained virtually the
same multiplicity ratios as with the virtuality-ordered shower. This
confirms the decoupling of the flavour and the fragmentation parameters
and no re-tuning of the flavour parameters with the \pT{}-ordered shower
is needed.

\subsection{Fragmentation optimisation}
\label{sec:fragmentationtune}

Based on the new flavour parameter settings the non-flavour
hadronisation and shower parameters were tuned, separately for the
virtuality-ordered and for the \pT{}-ordered shower. The observables
used in this step of the tuning were event shape variables, momentum
spectra, and the mean charged multiplicity measured by the \delphi{}
collaboration~\cite{Abreu:1996na}, momentum spectra and flavour-specific
mean charged multiplicities measured by the \opal{}
collaboration~\cite{Ackerstaff:1998hz}, and the $b$-quark fragmentation
function measured by the \delphi{} collaboration~\cite{delphi-2002}.

We tuned the same set of parameters for both shower types
(\Tab{tab:tune-frag}). To turn on the \pT{}-ordered shower, \MSTJ{41}
was set to 12 -- in the case of the virtuality-ordered shower, this
parameter stayed at its default value. For both tunes, we generated
1M~events at each of 100~parameter points.

During the tuning of the \pT{}-ordered shower it transpired that the fit
prefers uncomfortably low values of the shower cut-off \PARJ{82}. Since
this value needs to be at least $2 \cdot \Lambda_\text{QCD}$, and
preferably higher, it was manually fixed to 0.8 to keep the parameters
in a physically meaningful regime. Then the fit was repeated with the
remaining five parameters.

The second issue we encountered with the \pT{}-ordered shower was that
the polynomial parameterisation $f^{(b)}$ for the mean charged
multiplicity differed from the real Monte Carlo response by about 0.2
particles. This discrepancy was accounted for during the $\chi^2$
minimisation, so that the final result does not suffer from a bias in
this observable.

\begin{figure}
\begin{center}
\includegraphics[width=0.49\textwidth]{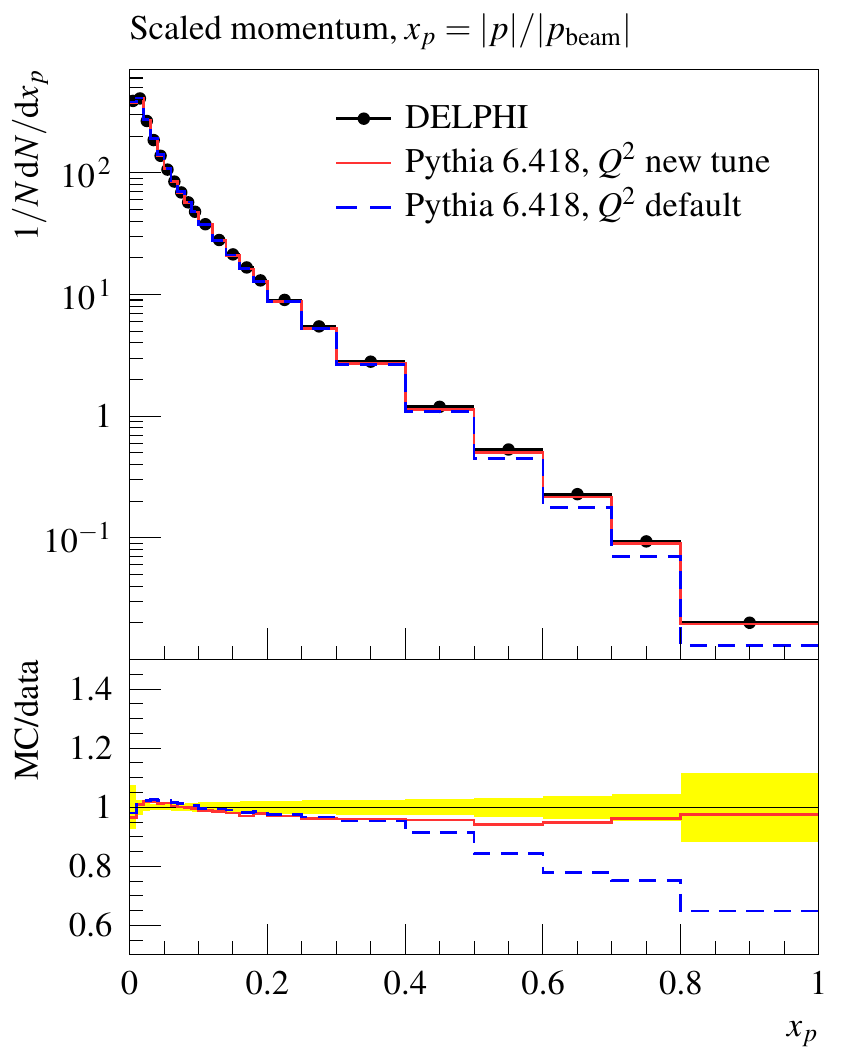}
\includegraphics[width=0.49\textwidth]{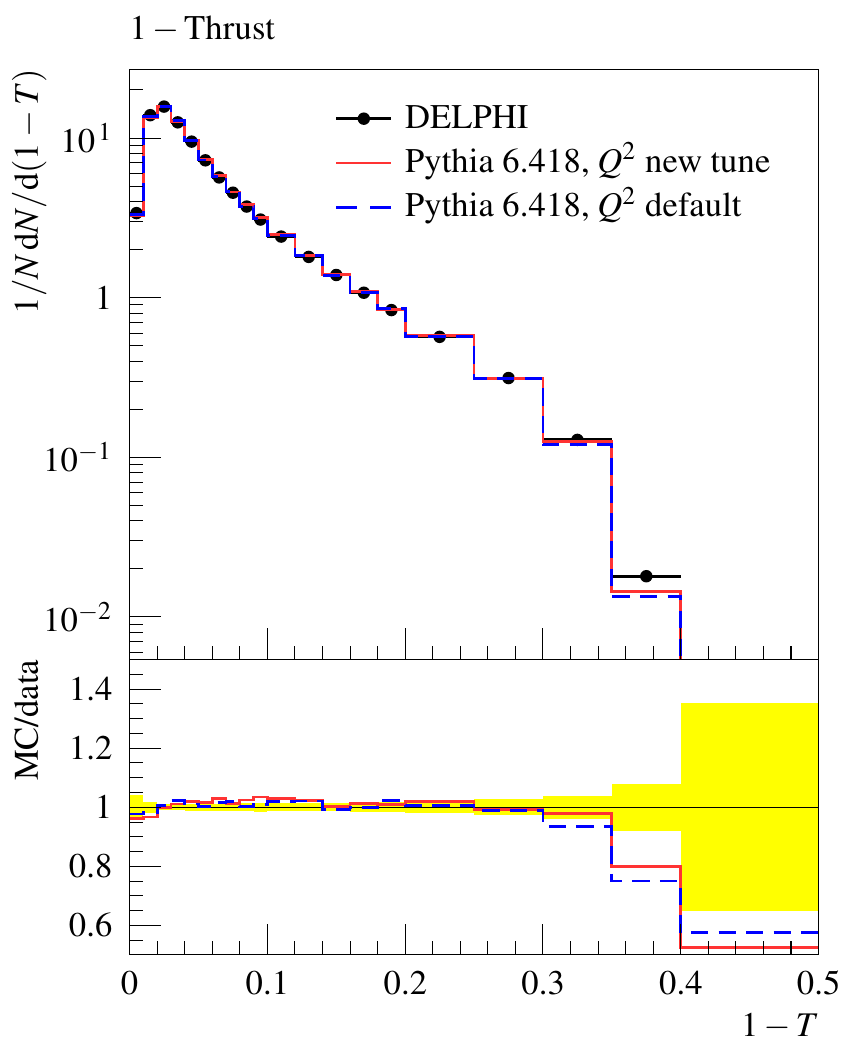}
\includegraphics[width=0.49\textwidth]{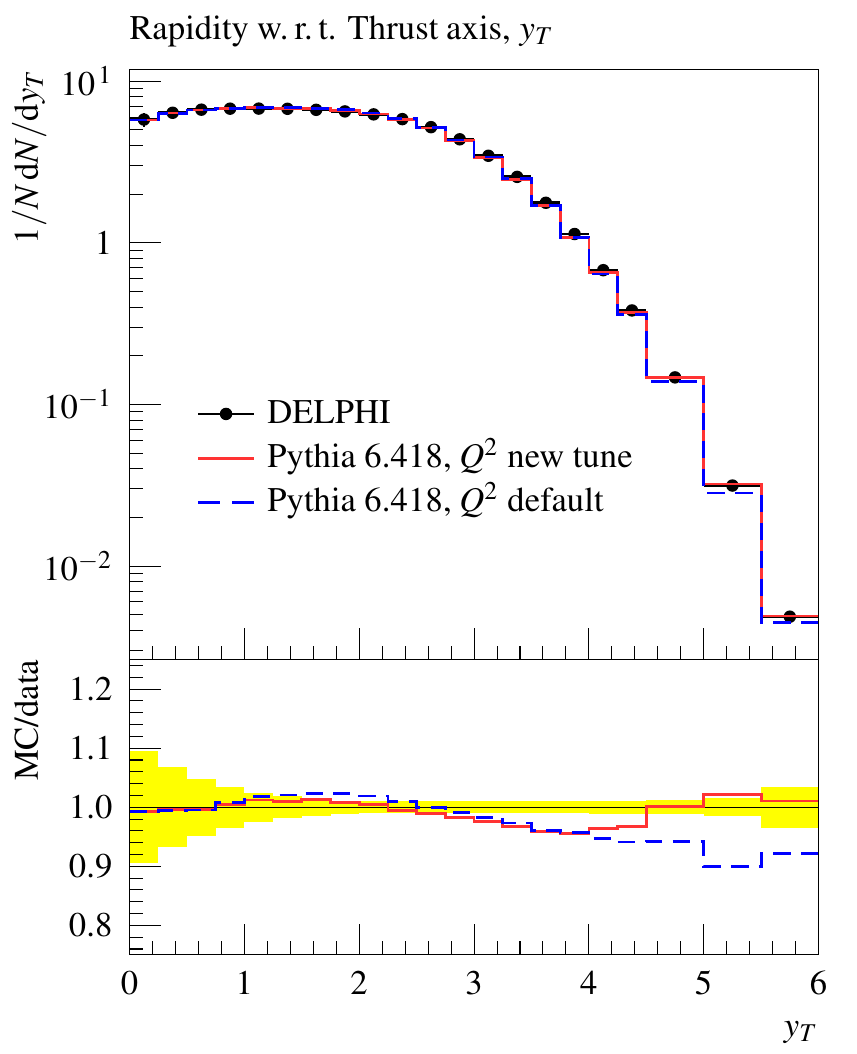}
\includegraphics[width=0.49\textwidth]{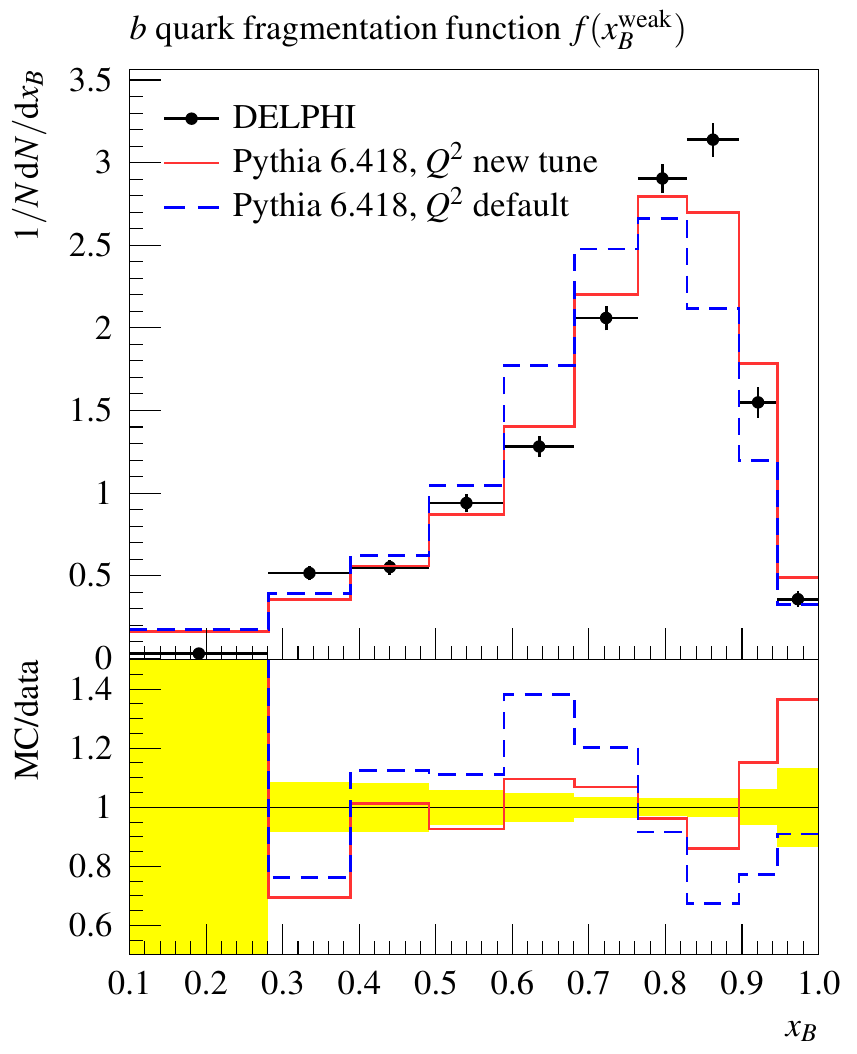}
\end{center}
\caption{Some example distributions for $e^+e^-$ collisions using the
         virtuality-ordered shower. The solid line shows the new tune,
         the dashed line is the default. Even though the
         virtuality-ordered shower is well-tested and \pythia{} has been
         tuned several times, especially by the \lep{} collaborations,
         there is still room for improvement in the default settings.
         Note the different scale in the ratio plot of the rapidity
         distribution.  The data in these plots has been published by
         \delphi{}~\cite{Abreu:1996na,delphi-2002}.}
\label{fig:tune-frag-q2}
\end{figure}

In \Fig{fig:tune-frag-q2} some comparison plots between the \pythia{}
default and our new tune of the virtuality-ordered shower are depicted.
Even though this shower has been around for many years and \pythia{} has
been tuned before, there still is room for improvement in the default
settings.

\begin{figure}
\begin{center}
\includegraphics[width=0.49\textwidth]{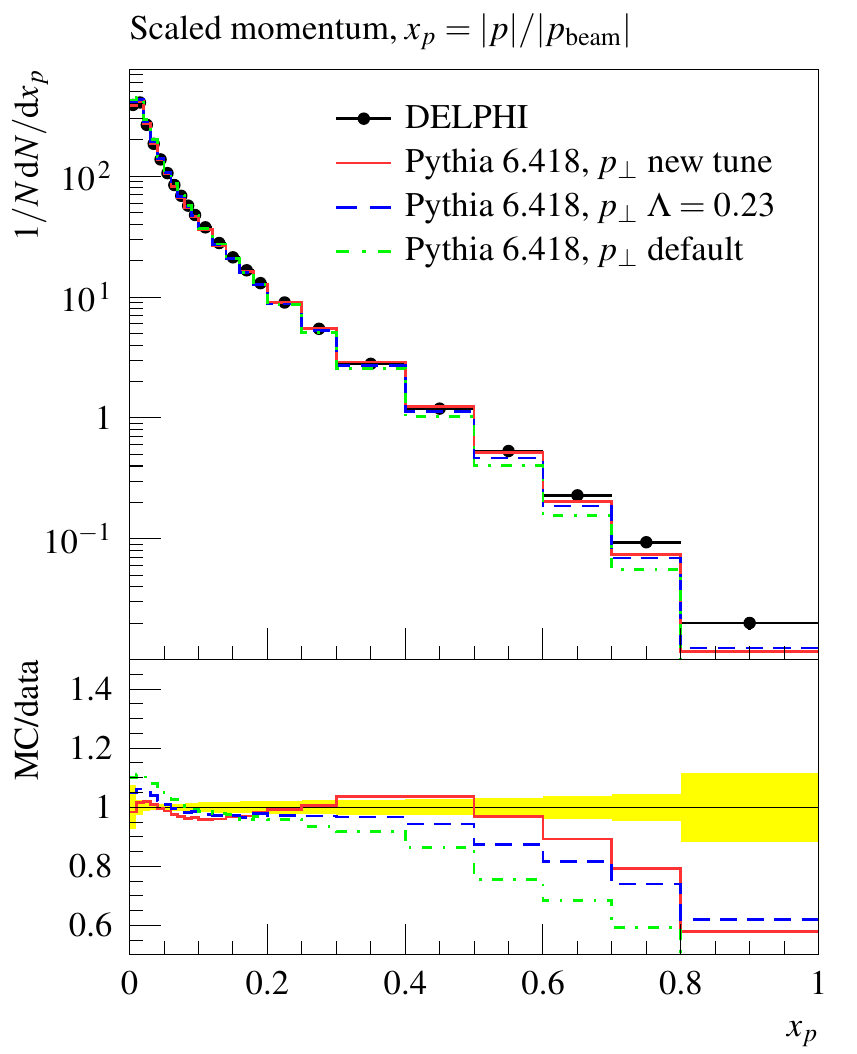}
\includegraphics[width=0.49\textwidth]{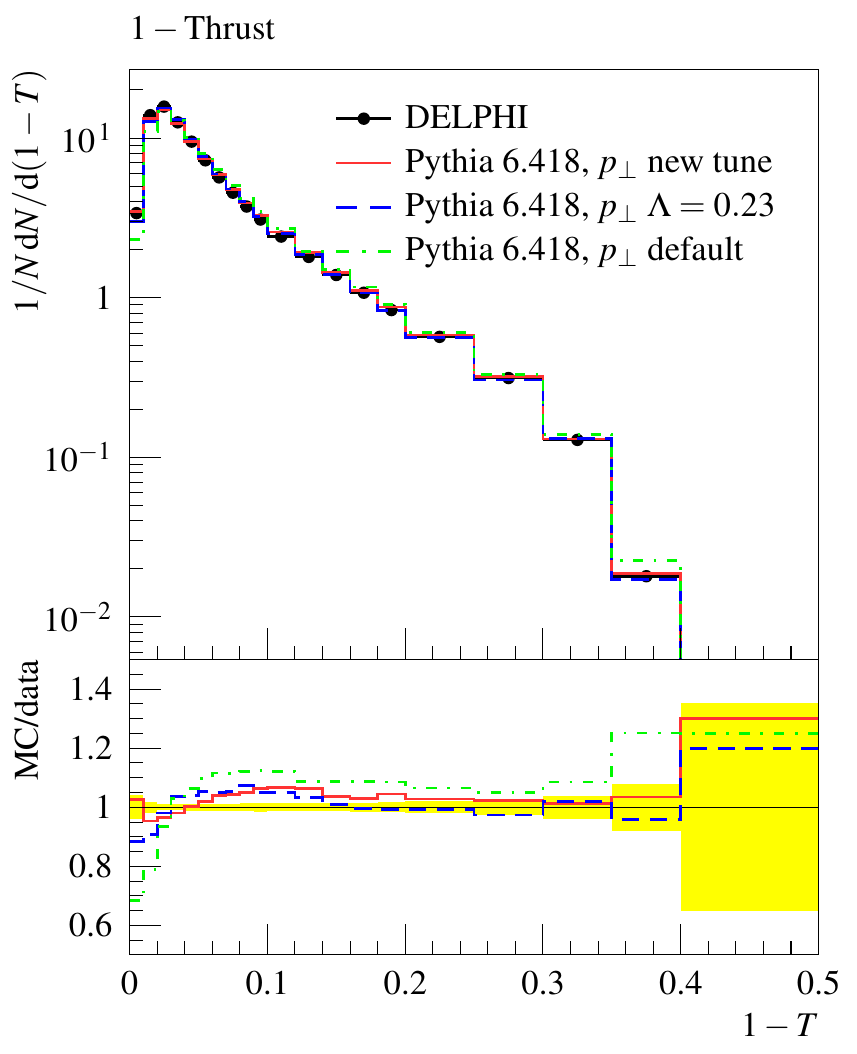}
\includegraphics[width=0.49\textwidth]{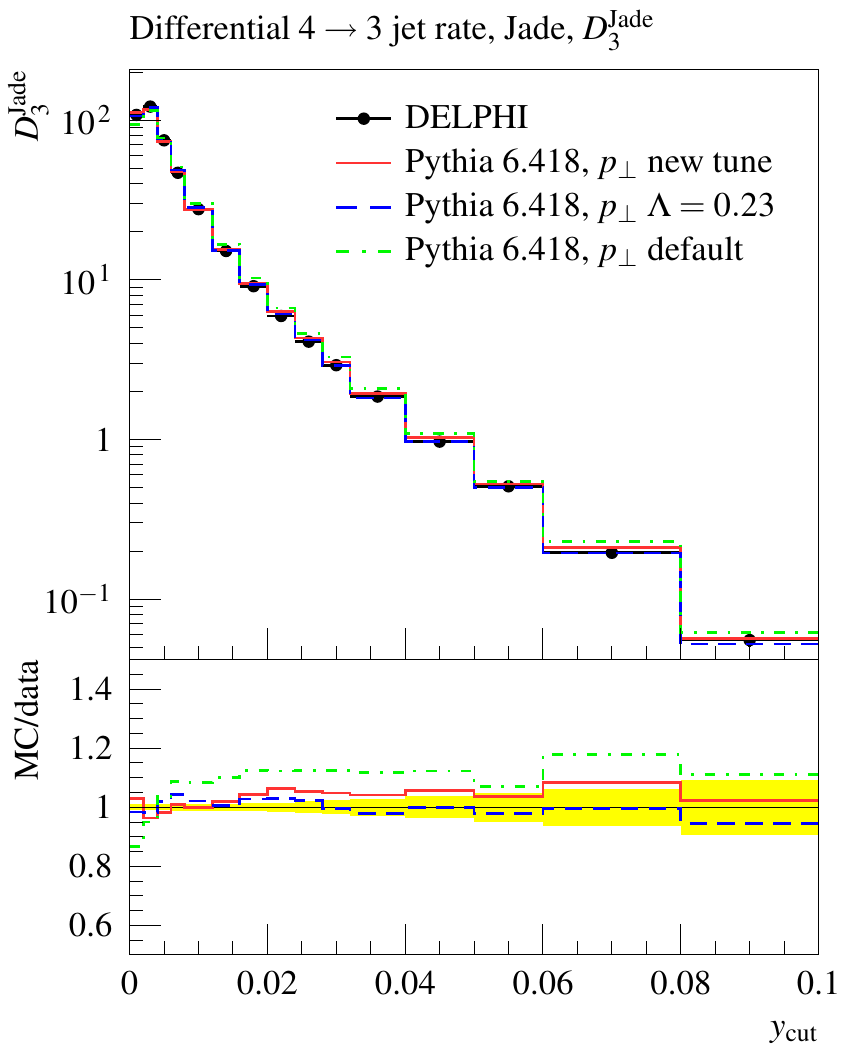}
\includegraphics[width=0.49\textwidth]{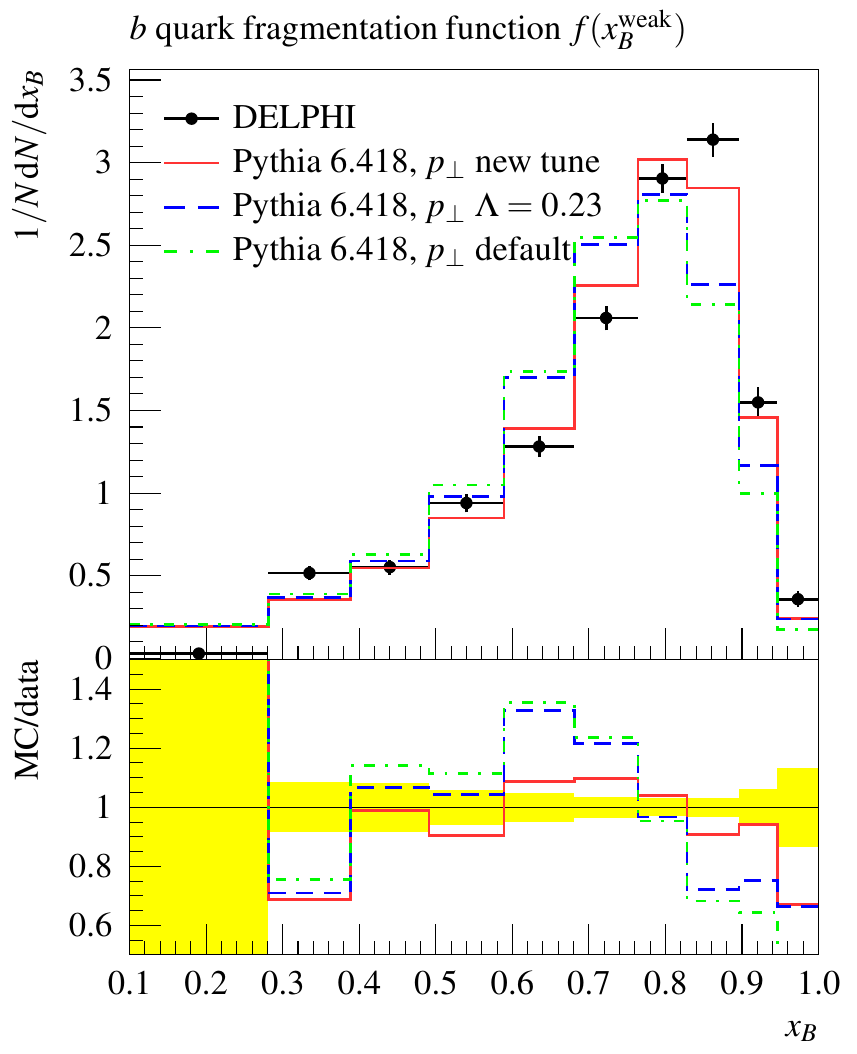}
\end{center}
\caption{Some example distributions for $e^+e^-$ collisions using the
         \pT{}-ordered shower. The solid line shows the new tune, the
         dashed line is the old recommendation for using the
         \pT{}-ordered shower (i.\,e. changing $\Lambda_\text{QCD}$ to
         0.23), and the dashed-dotted line is produced by switching on
         the \pT{}-ordered shower leaving everything else at its
         default.  The latter is the unfortunate choice made for the
         ATLAS-tune.  The data has been published by
         \delphi{}~\cite{Abreu:1996na,delphi-2002}.}
\label{fig:tune-frag-pt}
\end{figure}

\Fig{fig:tune-frag-pt} shows comparisons of the \pT{}-ordered shower.
This shower is a new option in \pythia{} and has not been tuned
systematically before. Nevertheless, the \pythia{} manual recommends to
set $\Lambda_\text{QCD}$ to 0.23. This recommendation is ignored by the
ATLAS collaboration, so our plots show our new tune, the default with
$\Lambda_\text{QCD}=0.23$, and the settings currently used by
ATLAS~\cite{moraes-tuneATLAS}.

\begin{table}
\begin{center}
\begin{tabular}{lrrrl}
\toprule
Parameter & Pythia 6.418 default & Final tune ($Q^2$) & Final tune ($\pT$) & \\
\midrule
\MSTJ{11}  & 4    & 5     & 5     & frag. function         \\
\PARJ{21}  & 0.36 & 0.325 & 0.313 & $\sigma_q$             \\
\PARJ{41}  & 0.3  & 0.5   & 0.49  & $a$                    \\
\PARJ{42}  & 0.58 & 0.6   & 1.2   & $b$                    \\
\PARJ{47}  & 1    & 0.67  & 1.0   & $r_b$                  \\
\PARJ{81}  & 0.29 & 0.29  & 0.257 & $\Lambda_\text{QCD}$   \\
\PARJ{82}  & 1    & 1.65  & 0.8   & shower cut-off         \\
\bottomrule
\end{tabular}
\end{center}
\caption{Tuned fragmentation parameters and their defaults for the
virtuality and \pT{}-ordered showers.}
\label{tab:tune-frag}
\end{table}

\subsection{Underlying event and multiple parton interactions}

For the third step we tuned the parameters relevant to the underlying event,
again both for the virtuality-ordered shower and the old MPI model, and
for the \pT{}-ordered shower with the interleaved MPI model.  This was
based on various Drell-Yan, jet physics, and minimum bias measurements
performed by \cdf{} and \dzero{} in \runi{} and
\runii{}~\cite{Affolder:1999jh,Affolder:2001xt,Acosta:2001rm,%
  Aaltonen:2009ne,cdf-note9351,cdf-leadingjet,Abazov:2004hm}.

The new MPI model differs significantly from the old one, hence we had
to tune different sets of parameters for these two cases. For the
virtuality-ordered shower and old MPI model we took Rick Field's
tune~DW~\cite{field-tuneDW} as guideline. In the case of the new model
we consulted Peter Skands and used a setup similar to his
tune~S0~\cite{Sandhoff:2005jh,Skands:2007zg} as starting point. All
switches and parameters for the UE/MPI tune, and our results are listed
in \Tabs{tab:params-ueq2} and~\ref{tab:params-uept}.

One of the main differences we observed between the models is their
behaviour in Drell-Yan physics. The old model had a hard time describing
the $Z$-\pT{} spectrum~\cite{Affolder:1999jh} and we had to assign a
high weight to that observable in order to force the Monte Carlo to get
the peak region of the distribution right (note that this is the only
observable to which we assigned different weights for the tunes of the
old and the new MPI model). The new model on the other hand gets the
$Z$-\pT{} right almost out of the box, but underestimates the underlying
event activity in Drell-Yan events as measured in~\cite{cdf-note9351}.
The same behaviour can be observed in Peter Skands'
tunes~\cite{Skands:2009zm}. We are currently investigating this
issue.

Another (albeit smaller) difference shows in the hump of the turn-on in
many of the UE distributions in jet physics. This hump is described by
the new model, but mostly missing in the old model. Although the origin
of this hump is thought to be understood, the model differences
responsible for its presence/absence in the two Pythia models is not yet
known in any detail.

\begin{figure}
\begin{center}
\includegraphics[width=0.49\textwidth]{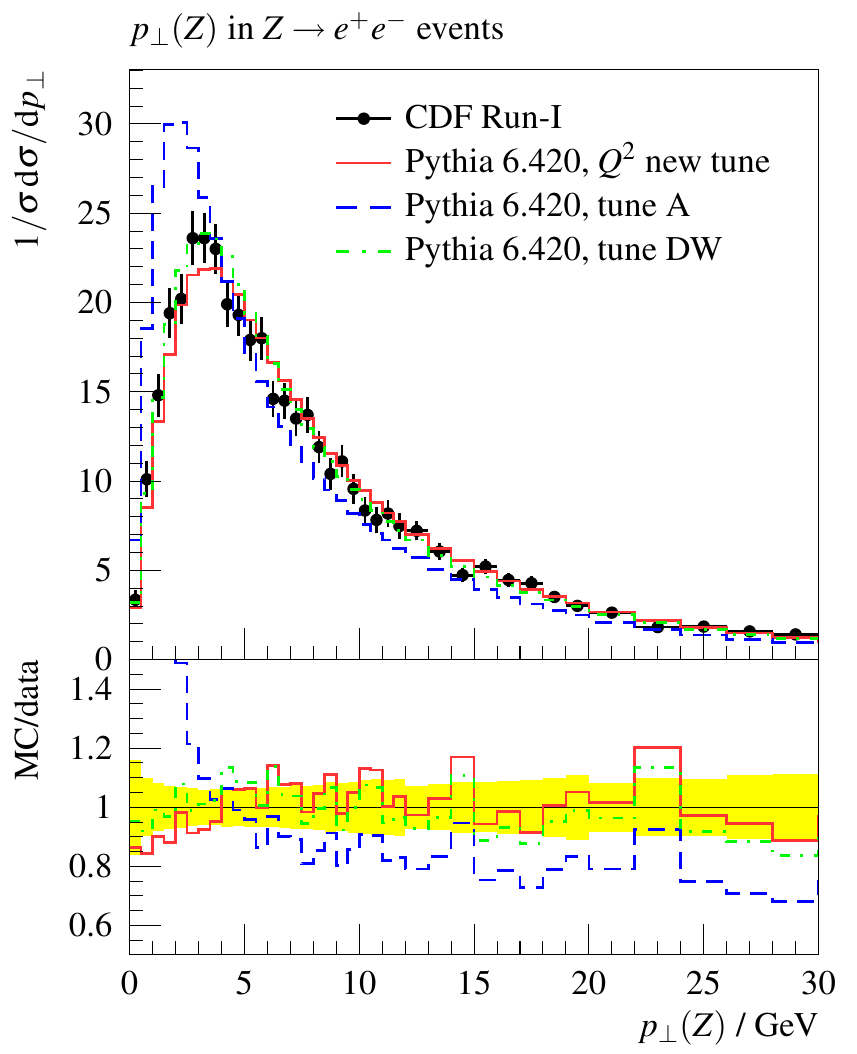}
\includegraphics[width=0.49\textwidth]{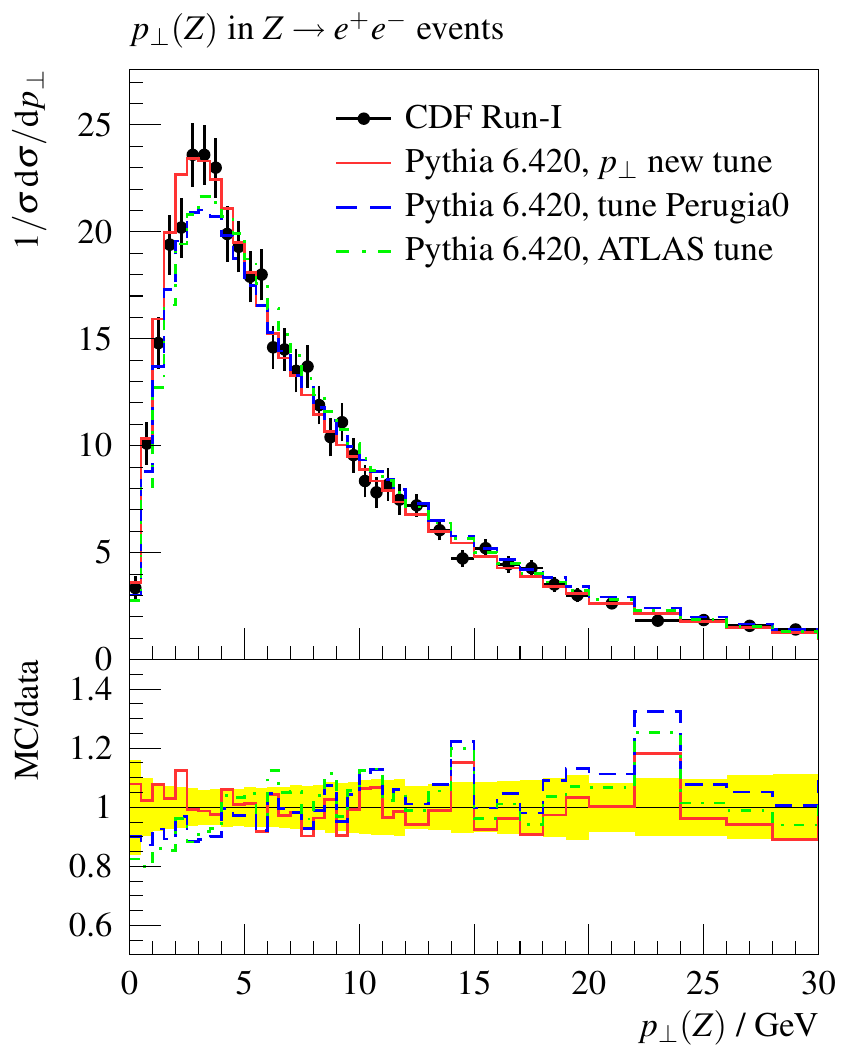}
\includegraphics[width=0.49\textwidth]{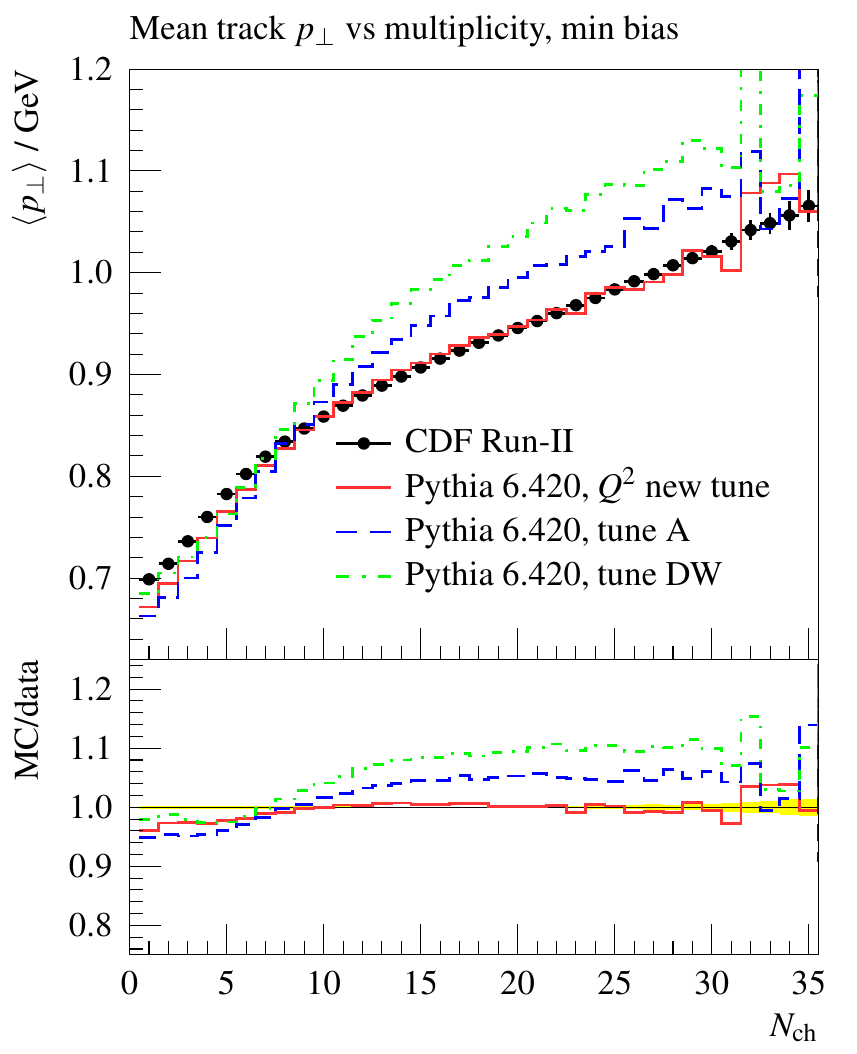}
\includegraphics[width=0.49\textwidth]{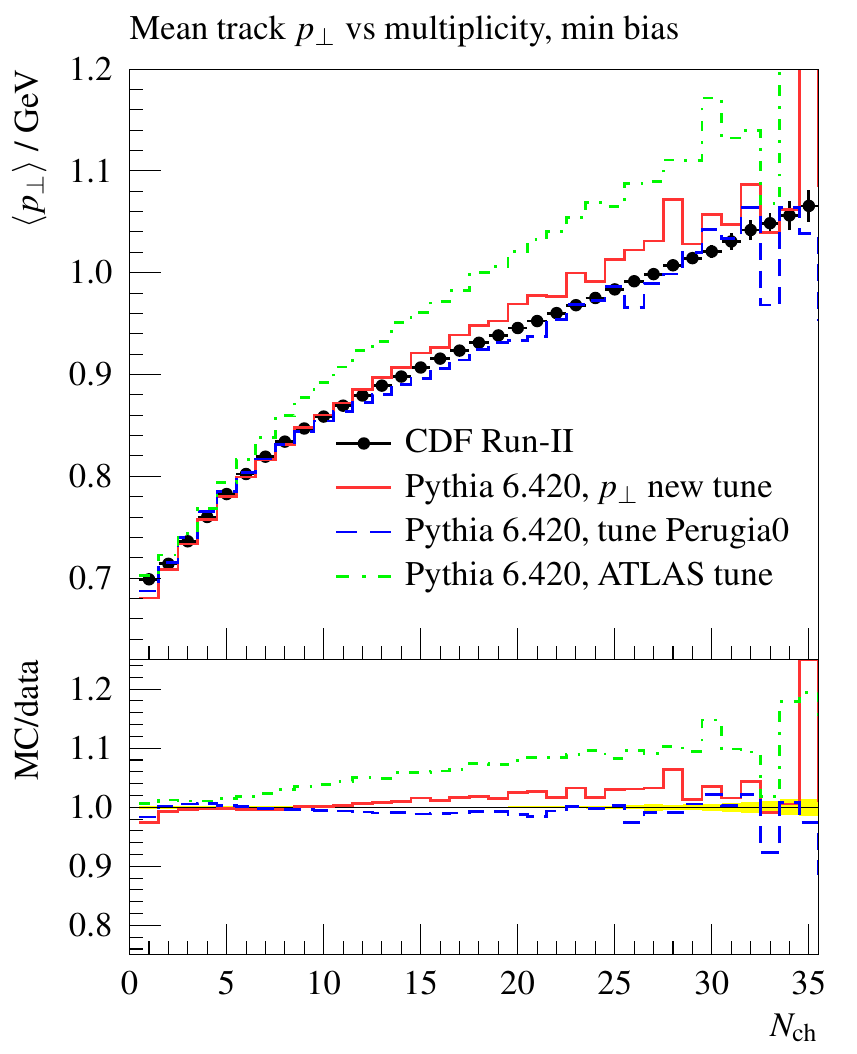}
\end{center}
\caption{The upper plots show the $Z$ \pT{} distribution as measured by
         CDF~\cite{Affolder:1999jh} compared to different tunes
         of the virtuality-ordered shower with the old MPI model (left)
         and the \pT{}-ordered shower with the interleaved MPI model
         (right). Except for tune~A all tunes describe this observable,
         and also the fixed version of tune~A, called AW, is basically
         identical to DW. The lower plots show the average track \pT{}
         as function of the charged multiplicity in minimum bias
         events~\cite{Aaltonen:2009ne}. This observable is quite
         sensitive to colour reconnection. Only the recent tunes hit the
         data here (except for ATLAS).}
\label{fig:tune-ue-1}
\end{figure}

\begin{figure}
\begin{center}
\includegraphics[width=0.49\textwidth]{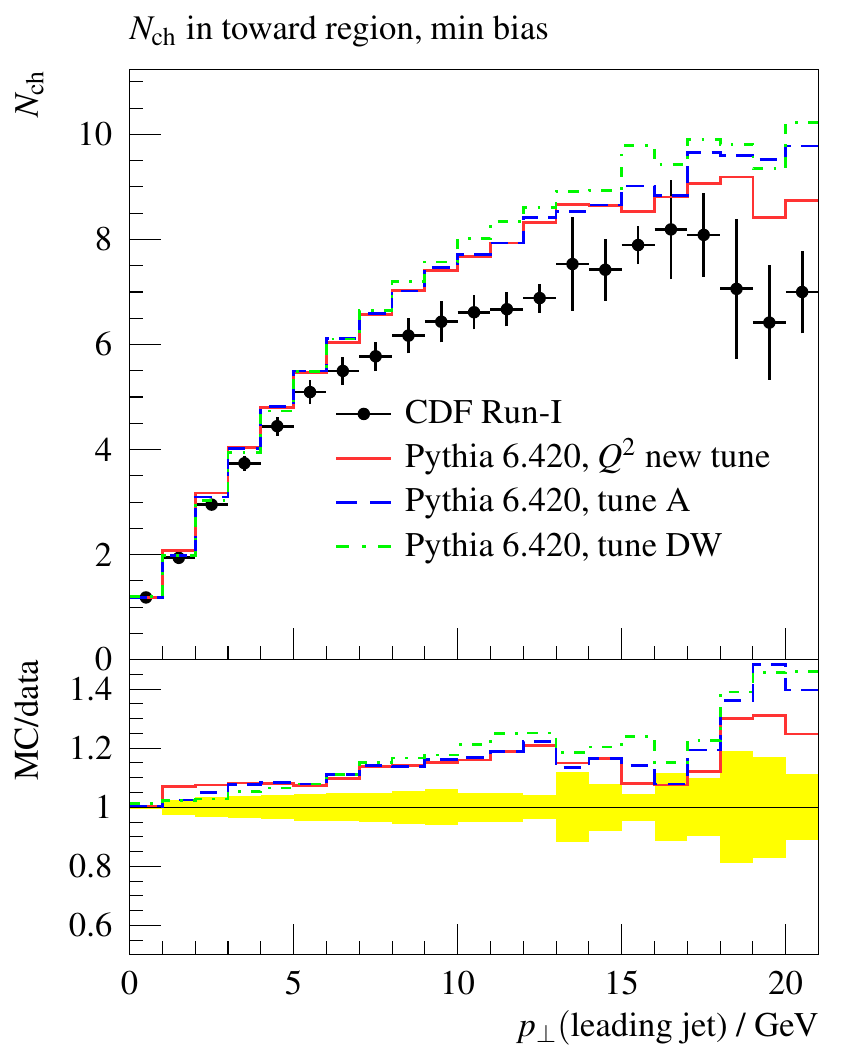}
\includegraphics[width=0.49\textwidth]{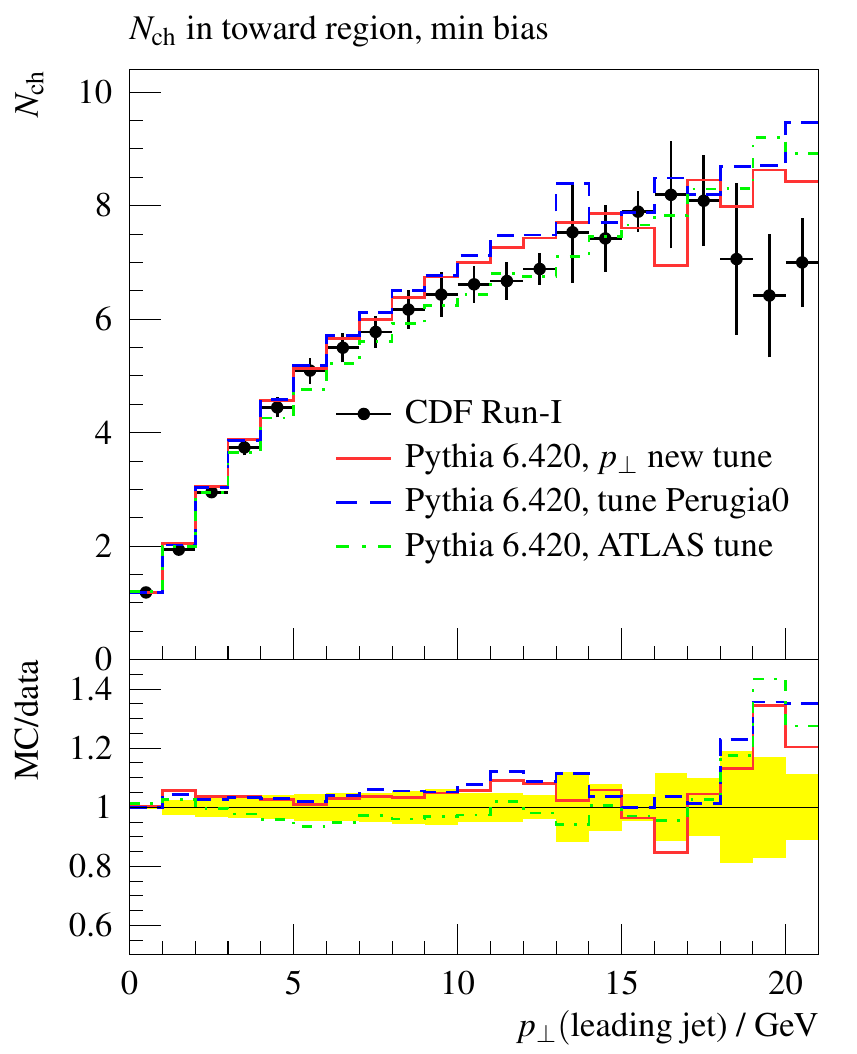}
\includegraphics[width=0.49\textwidth]{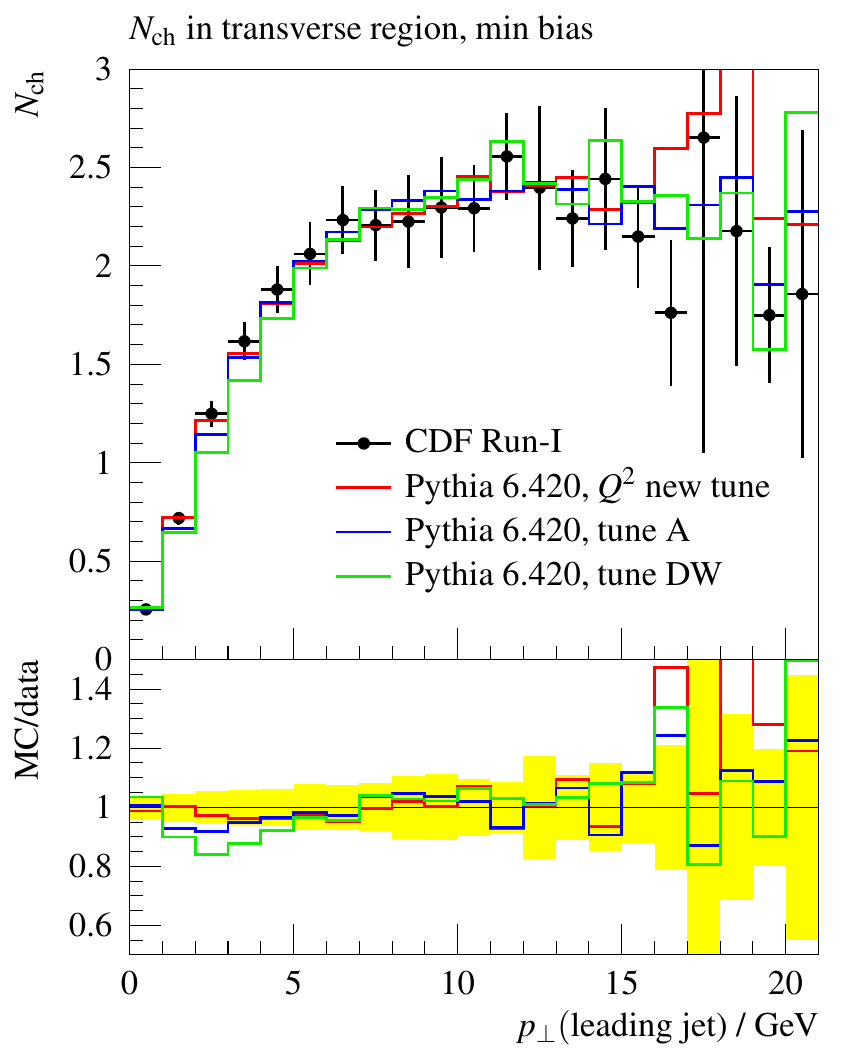}
\includegraphics[width=0.49\textwidth]{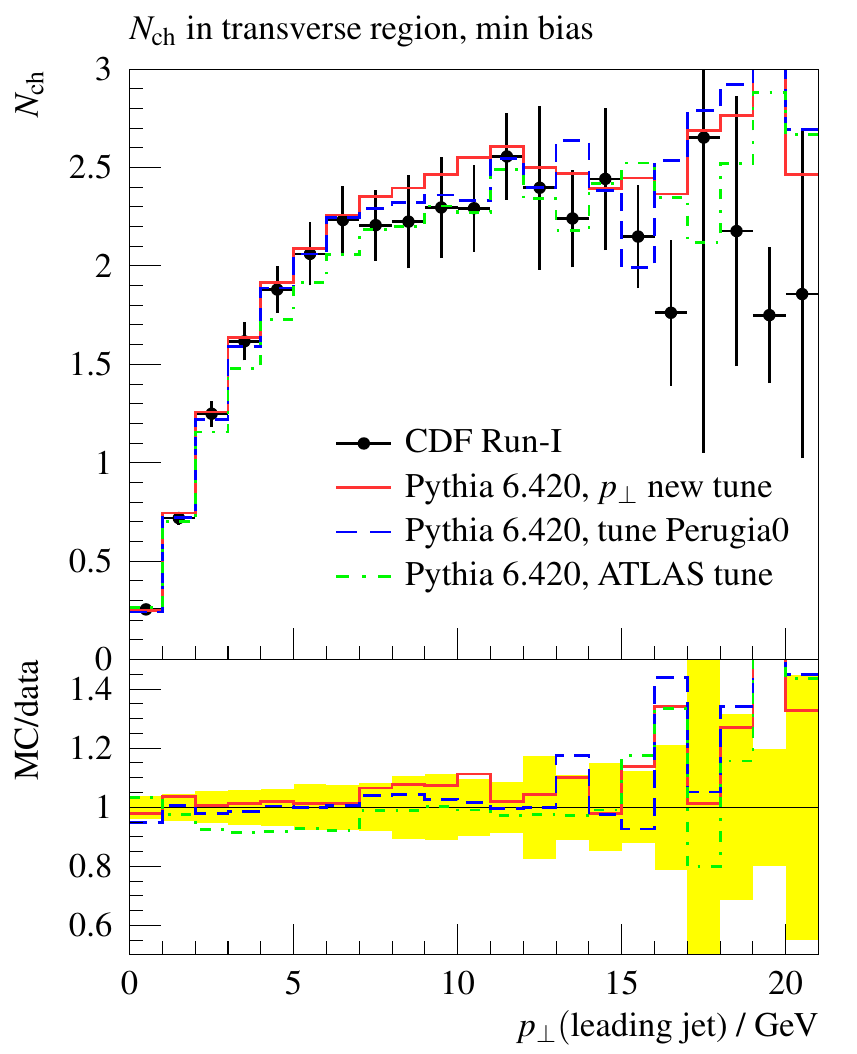}
\end{center}
\caption{These plots show the average charged multiplicity in the toward
         and transverse regions as function of the leading jet \pT{} in
         minimum bias events~\cite{Affolder:2001xt}. On the left side
         tunes of the virtuality-ordered shower with the old MPI model
         are shown, while on the right side the \pT{}-ordered shower
         with the interleaved MPI model is used. The old model is known
         to be a bit too ``jetty'' in the toward region, which can be
         seen in the first plot. Other than this, all tunes are very
         similar.}
\label{fig:tune-ue-2}
\end{figure}

\begin{figure}
\begin{center}
\includegraphics[width=0.49\textwidth]{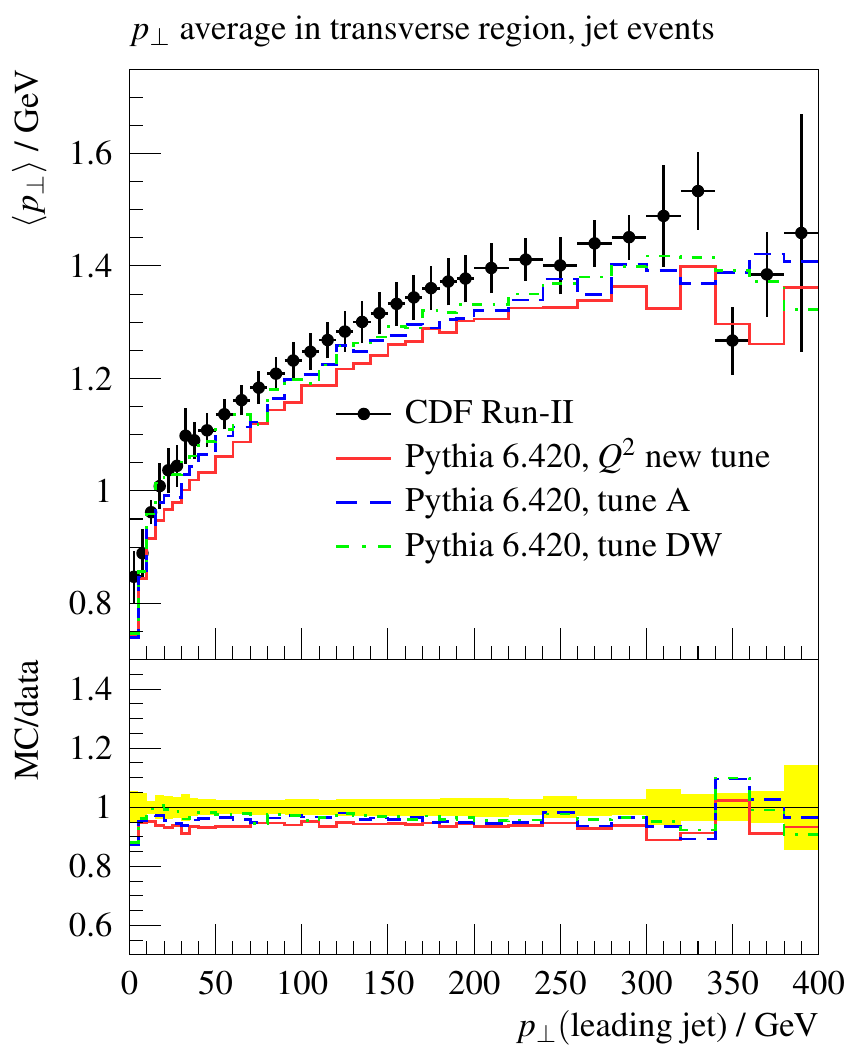}
\includegraphics[width=0.49\textwidth]{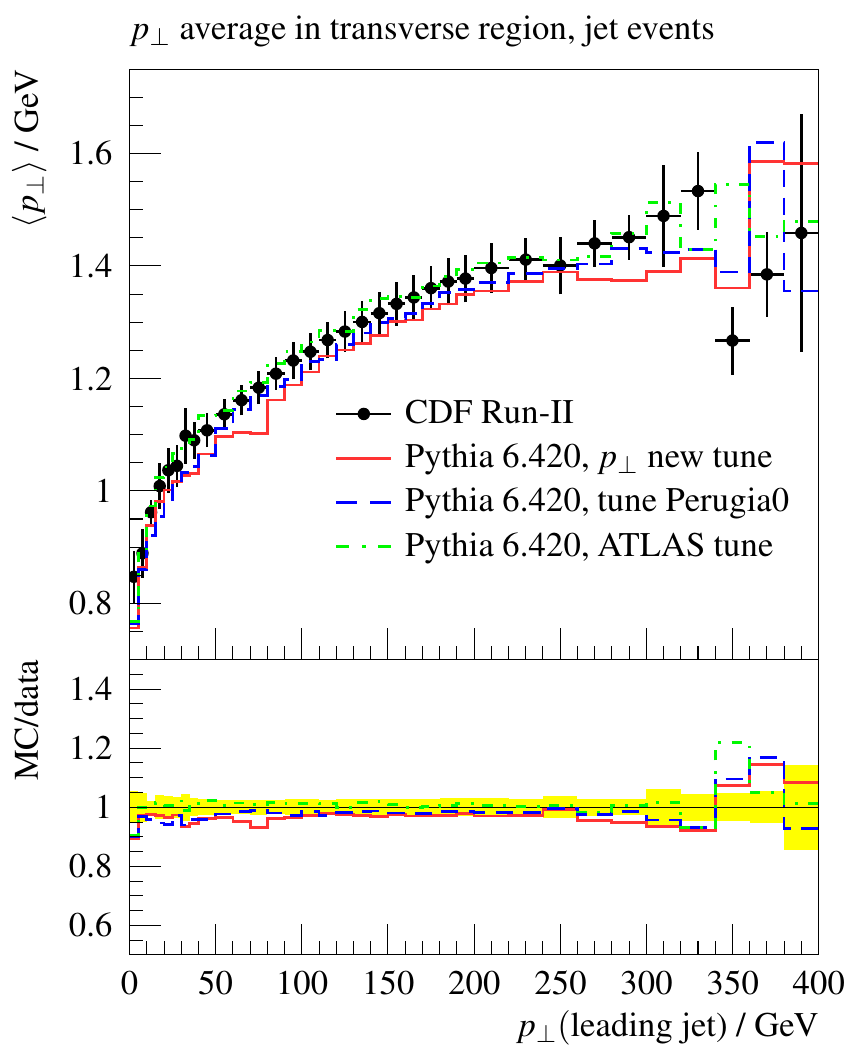}
\includegraphics[width=0.49\textwidth]{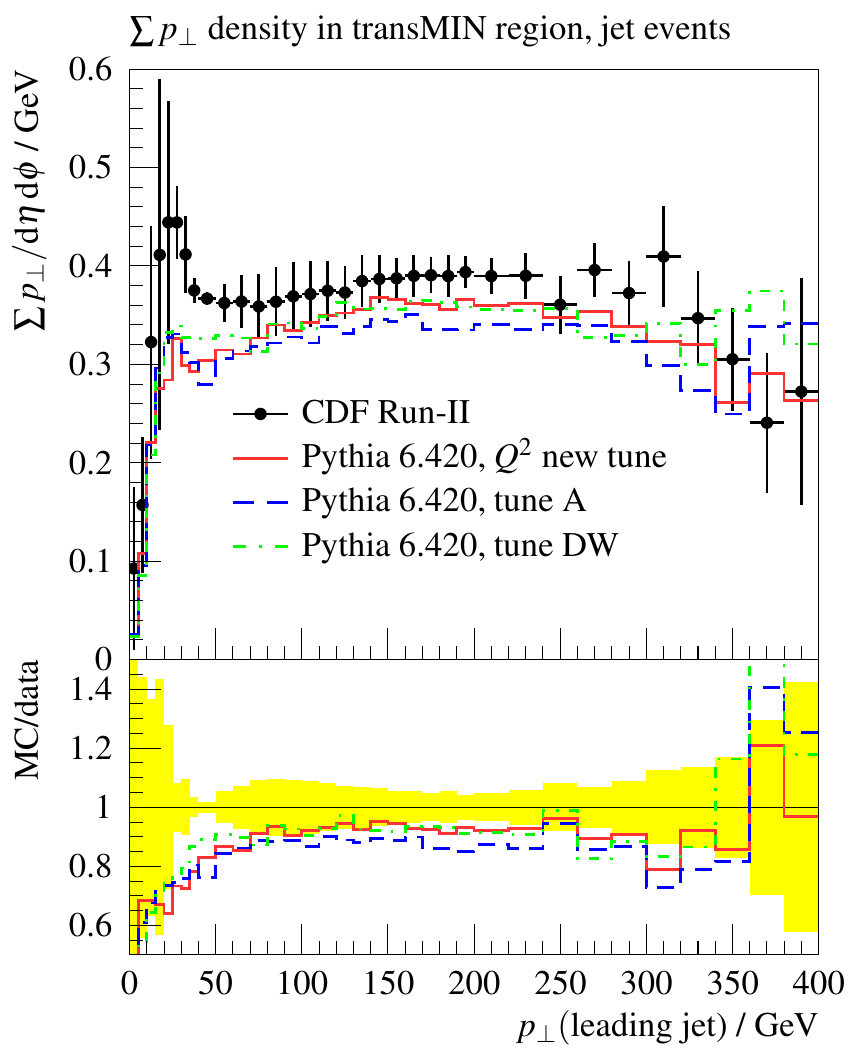}
\includegraphics[width=0.49\textwidth]{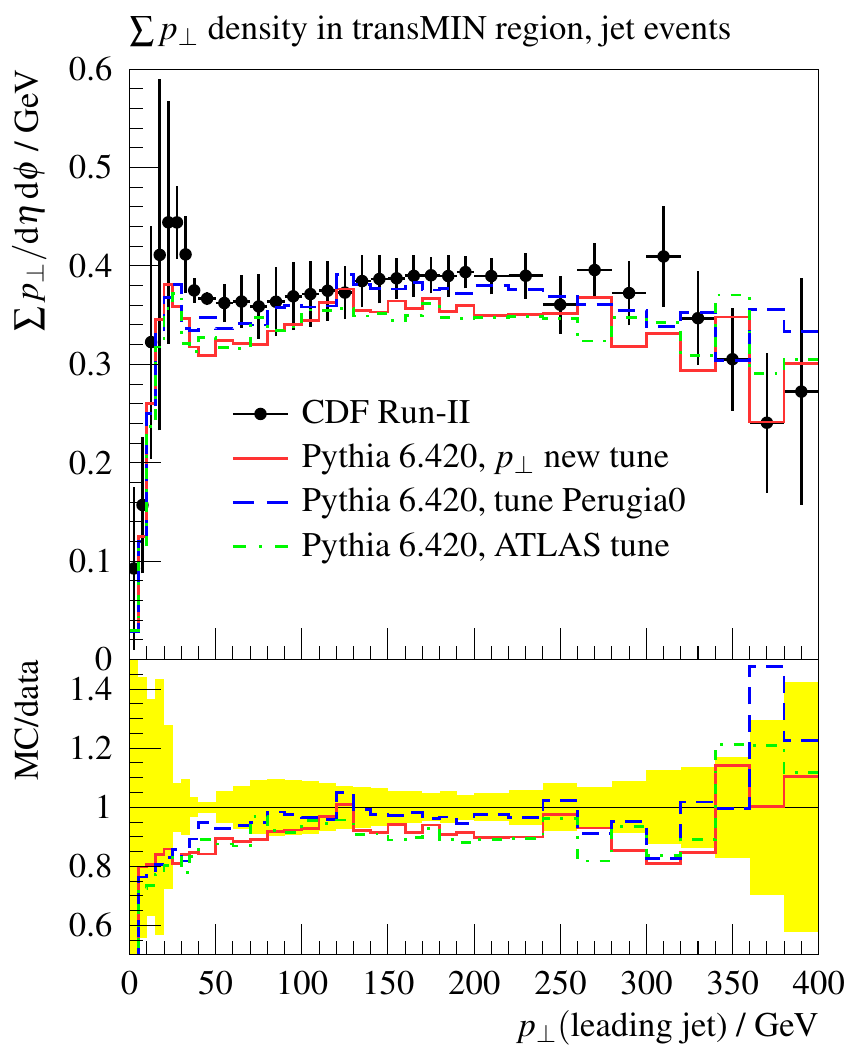}
\end{center}
\caption{These plots show the average track \pT{} in the transverse region
         (top) and the $\sum \pT$ density in the transMIN region
         (bottom) in leading jet events~\cite{cdf-leadingjet}. The new
         model (on the right) seems do have a slight advantage over the
         virtuality-ordered shower with the old MPI model shown on the left,
         both in the turn-on hump and in overall activity.}
\label{fig:tune-ue-3}
\end{figure}

\begin{figure}
\begin{center}
\includegraphics[width=0.49\textwidth]{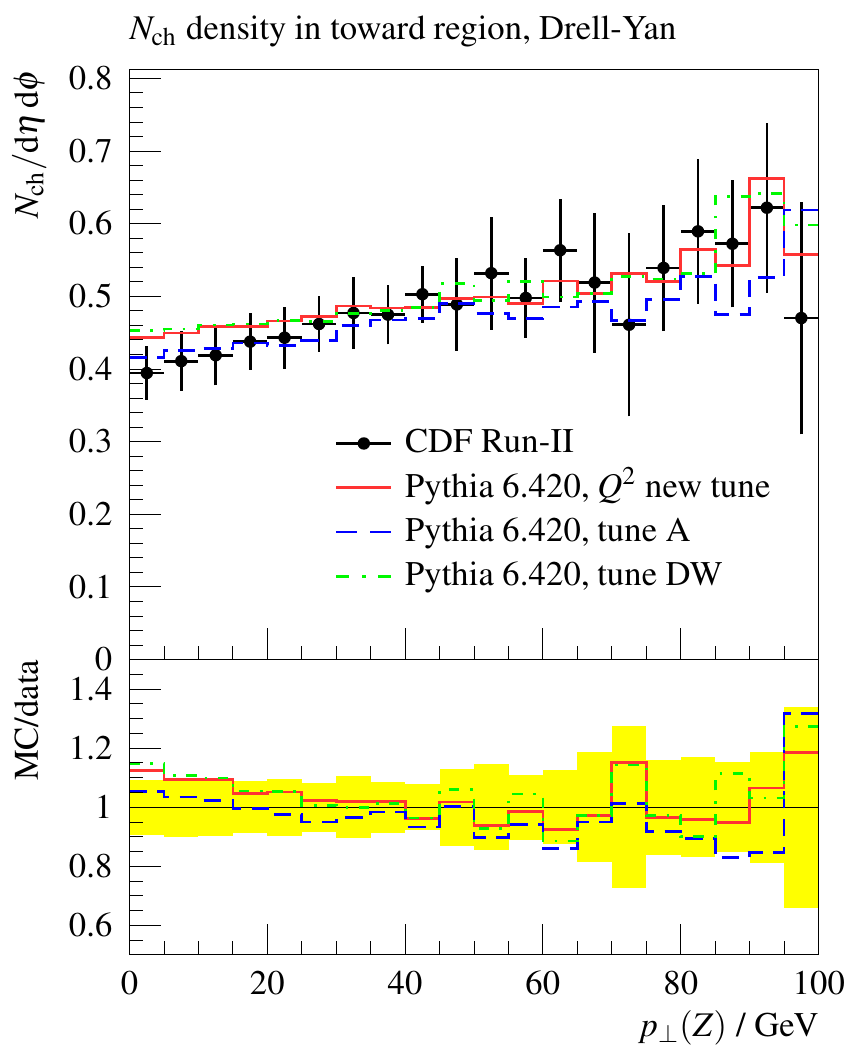}
\includegraphics[width=0.49\textwidth]{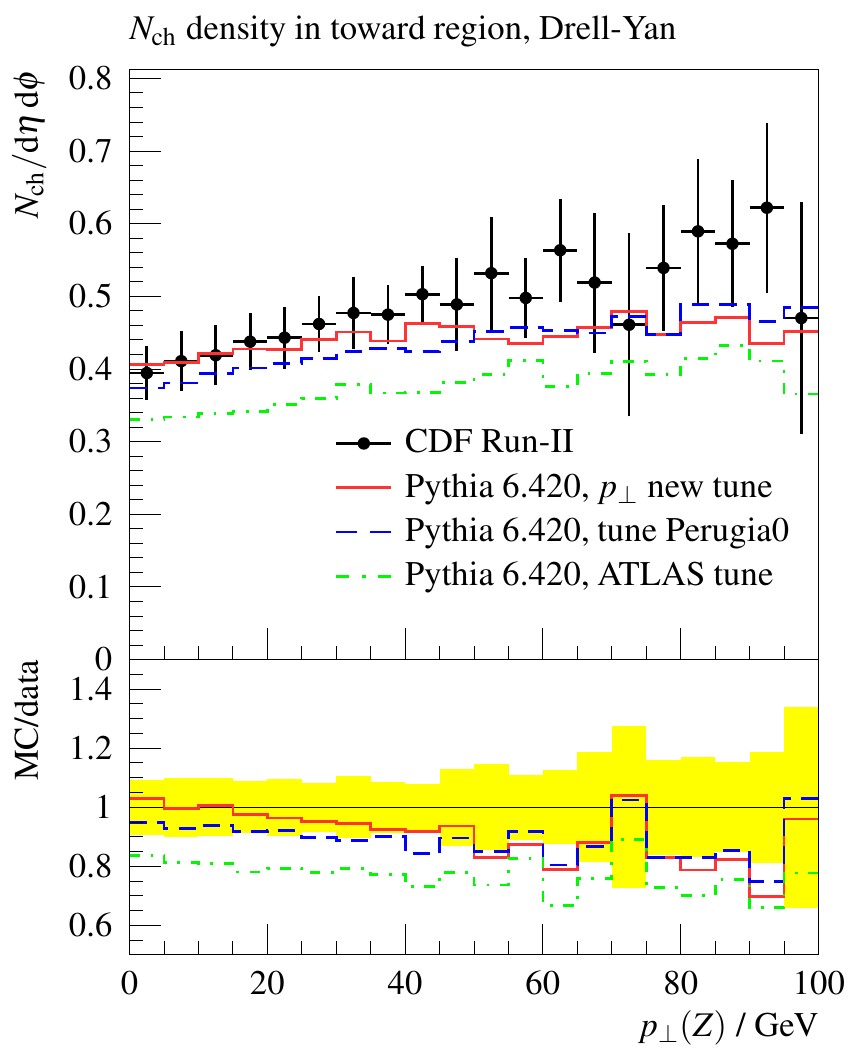}
\includegraphics[width=0.49\textwidth]{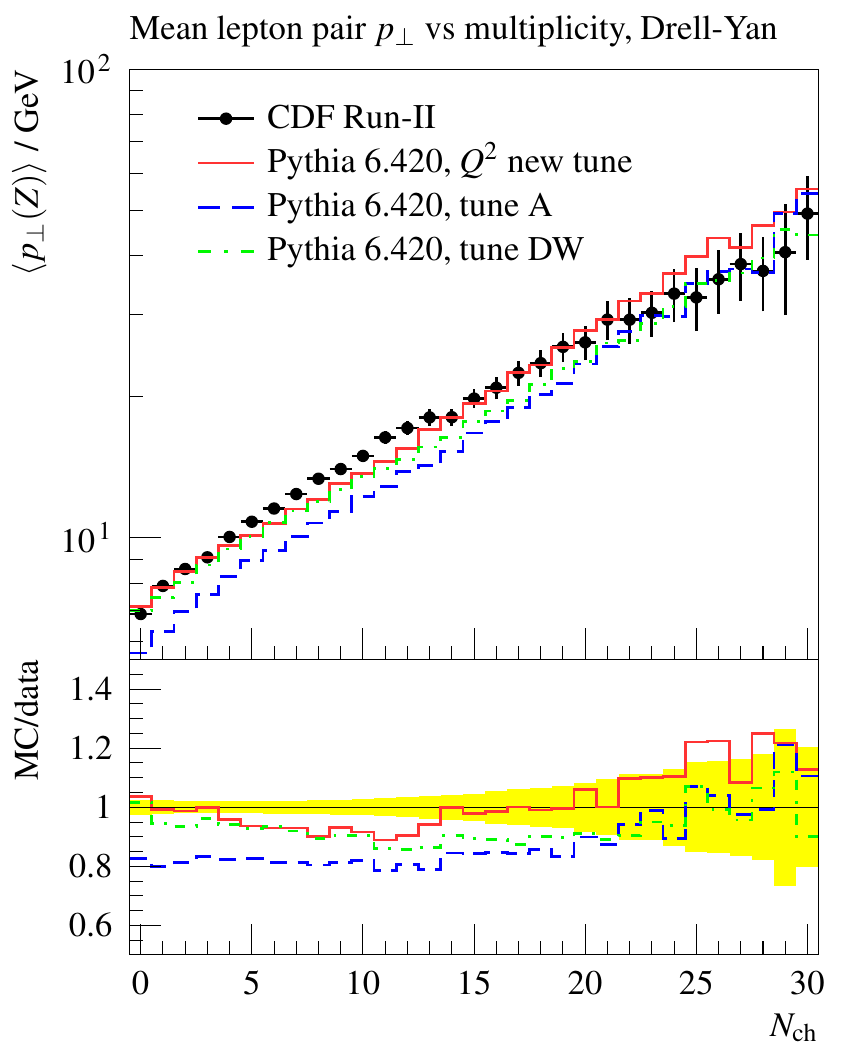}
\includegraphics[width=0.49\textwidth]{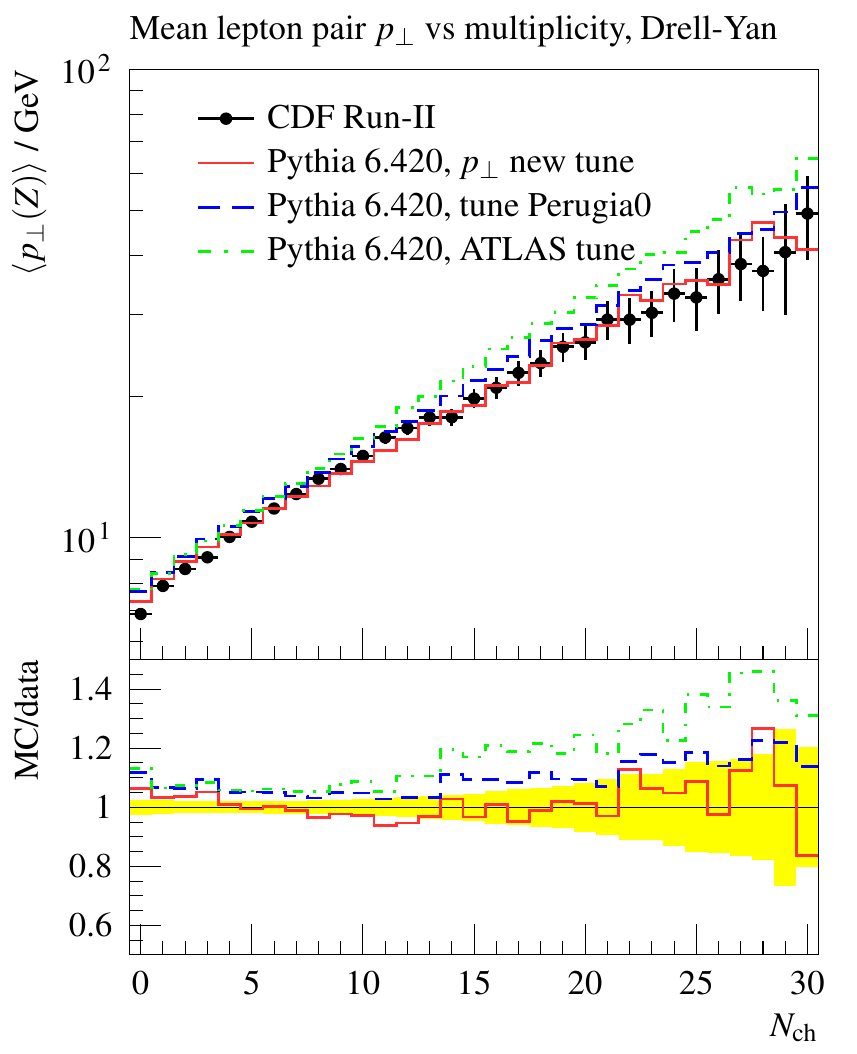}
\end{center}
\caption{In Drell-Yan~\cite{cdf-note9351} the new MPI model consistently
         underestimates the activity of the underlying event.
         Nevertheless, most of the recent tunes are able to describe the
         multiplicity dependence of the $Z$ \pT{}.}
\label{fig:tune-ue-4}
\end{figure}

\begin{figure}
\begin{center}
\includegraphics[width=0.49\textwidth]{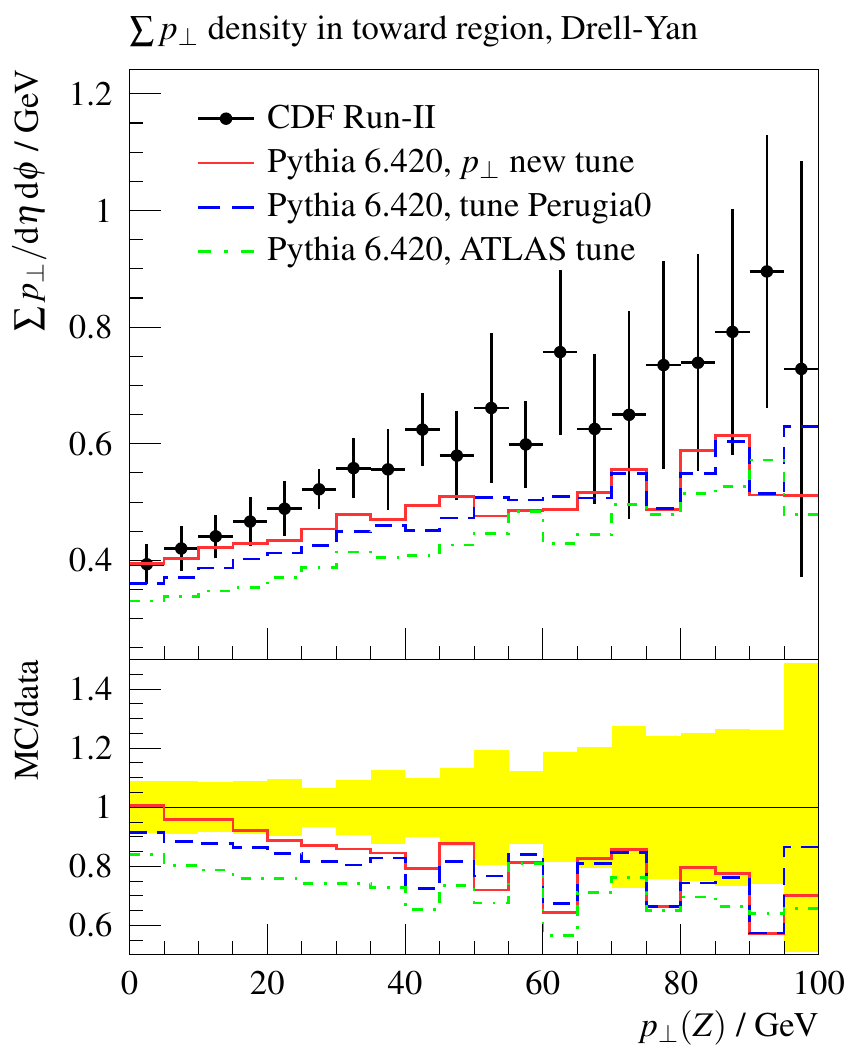}
\includegraphics[width=0.49\textwidth]{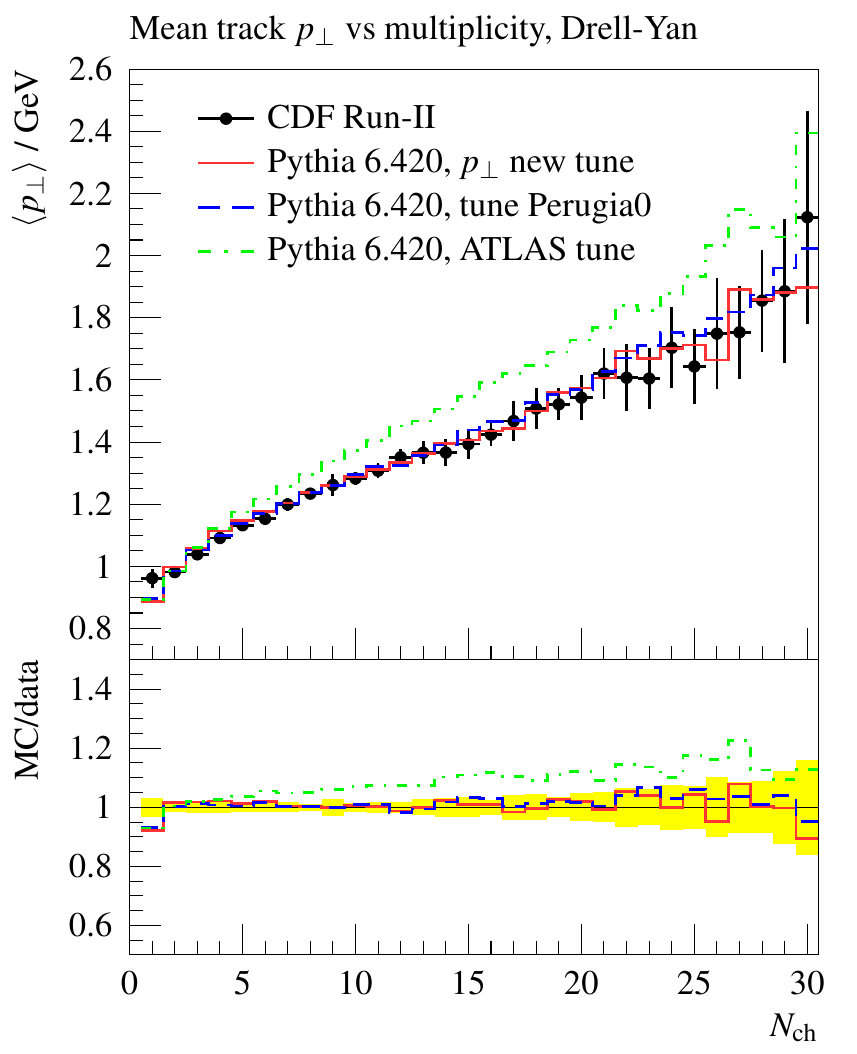}
\includegraphics[width=0.49\textwidth]{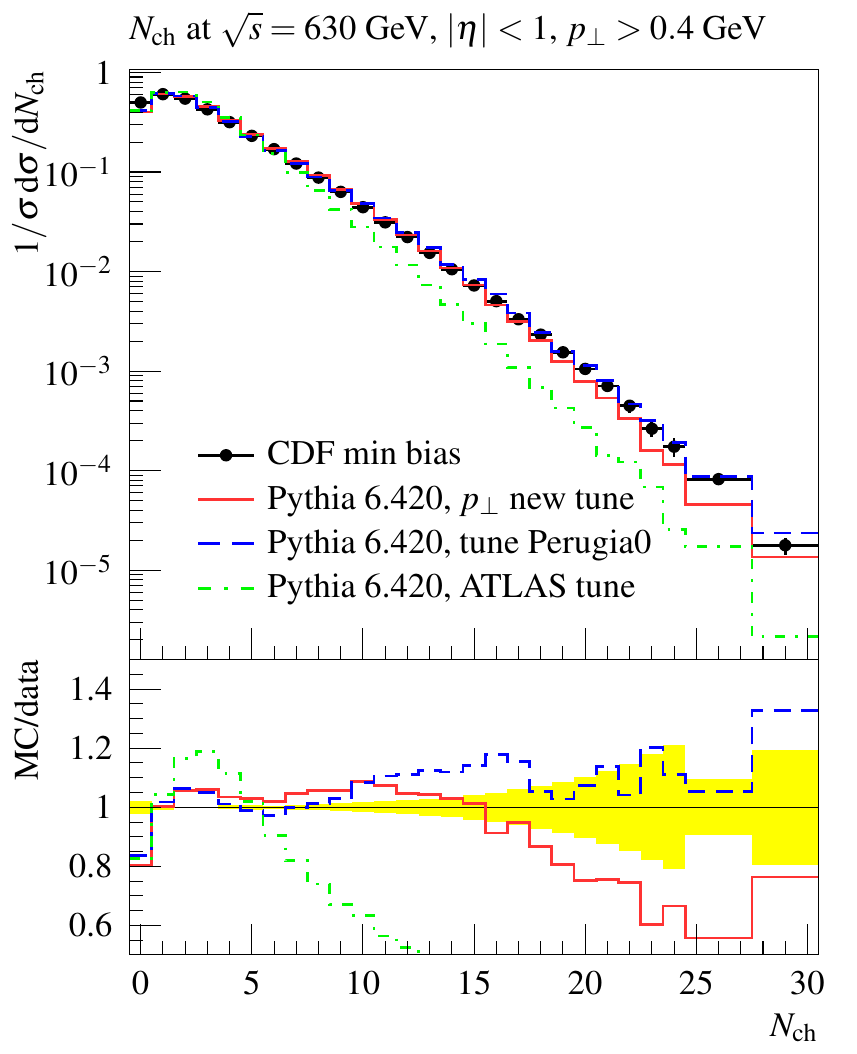}
\includegraphics[width=0.49\textwidth]{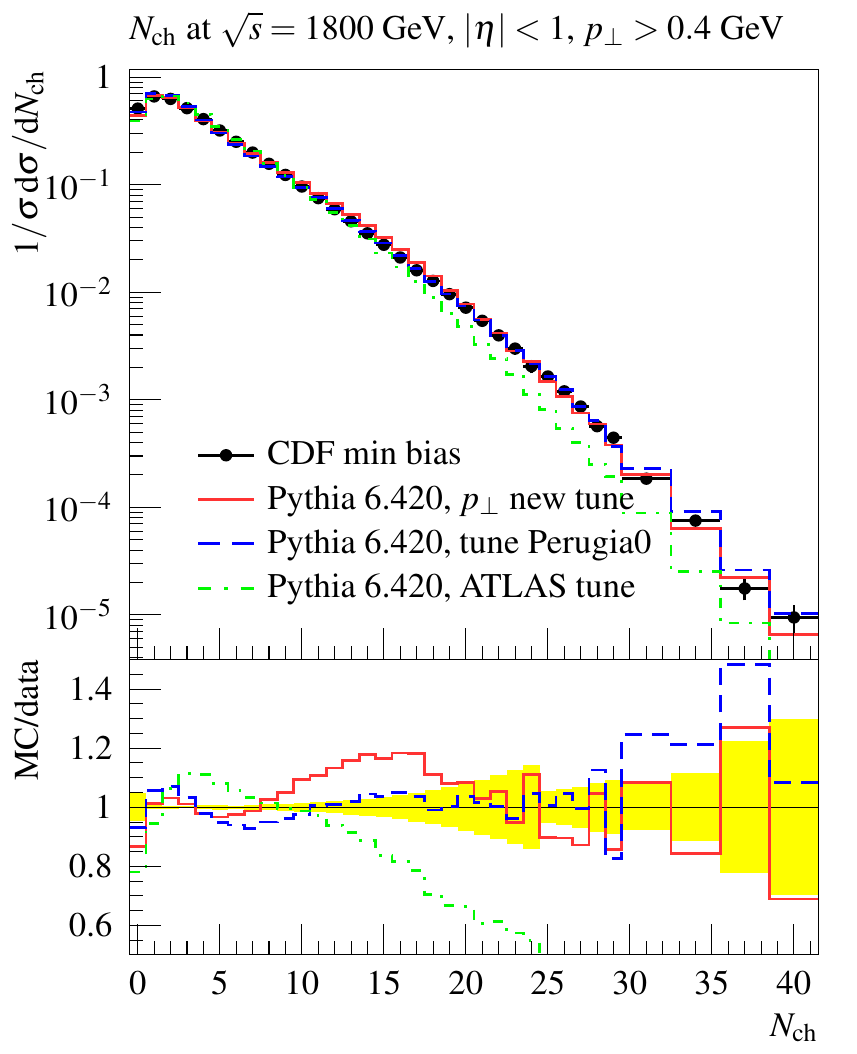}
\end{center}
\caption{Some more plots showing the behaviour of the interleaved MPI
         model and the \pT{}-ordered shower. The two upper plots focus
         on the underlying event in Drell-Yan~\cite{cdf-note9351}. On
         the left we see again that the new model underestimates the
         activity in Drell-Yan events (like in \Fig{fig:tune-ue-4}).
         Regardless of that, the top right plot shows that the average
         track \pT{} as function of the charged multiplicity is
         described well~-- except by the ATLAS tune. The ATLAS tune also
         has a big problem with the multiplicity distribution in minimum
         bias events shown in the lower two plots~\cite{Acosta:2001rm}.
         Even at the reference energy of 1800\,\gev{} this tune fails to
         match the data.}
\label{fig:tune-ue-5}
\end{figure}

\interfootnotelinepenalty=10000
\enlargethispage{\baselineskip}
\Figs{fig:tune-ue-1} to \ref{fig:tune-ue-5} show some comparisons
between our new tune and various other tunes.
For the virtuality-ordered shower with the old
MPI model we show Rick Field's tunes~A~\cite{field-tuneA} and
DW~\cite{field-tuneDW} as references, since they are well-known and
widely used. For the \pT{}-ordered shower and the new MPI framework we
compare to Peter Skands' new Perugia0 tune \cite{Skands:2009zm}. We
also include the current ATLAS tune~\cite{moraes-tuneATLAS} (even though
we don't believe it has good predictive power\footnote{Not only is the
choice of fragmentation parameters unfortunate (as discussed in
Section~\ref{sec:fragmentationtune}) and the tune fails to describe the
underlying event in Drell-Yan events, but also the energy scaling
behaviour in this tune is pretty much ruled out by the data
\cite{energyscaling}, making it in our eyes a particularly bad choice
for \lhc{} predictions.}), since it is widely used at the \lhc{}.

\begin{table}
\begin{center}
\begin{tabular}{lrrl}
\toprule
Parameter & \multicolumn{1}{l}{Pythia 6.418 default} & \multicolumn{1}{l}{Final tune} &  \\
\midrule
\PARP{62}  & 1.0 & 2.9 & ISR cut-off \\
\PARP{64}  & 1.0 & 0.14 & ISR scale factor for $\alpha_S$ \\
\PARP{67}  & 4.0 & 2.65 & max. virtuality \\
\PARP{82}  & 2.0 & 1.9 & $\pT^0$ at reference $E_\text{cm}$ \\
\PARP{83}  & 0.5 & 0.83 & matter distribution \\
\PARP{84}  & 0.4 & 0.6 & matter distribution \\
\PARP{85}  & 0.9 & 0.86 & colour connection \\
\PARP{86}  & 1.0 & 0.93 & colour connection \\
\PARP{90}  & 0.2 & 0.22 & $\pT^0$ energy evolution \\
\PARP{91}  & 2.0 & 2.1 & intrinsic $k_\perp$ \\
\PARP{93}  & 5.0 & 5.0 & intrinsic $k_\perp$ cut-off \\
\bottomrule
\end{tabular}
\end{center}
\caption{Tuned parameters for the underlying event using the virtuality-ordered shower}
\label{tab:params-ueq2}
\end{table}

\begin{table}
\begin{center}
\begin{tabular}{lrrl}
\toprule
Parameter & \multicolumn{1}{l}{Pythia 6.418 default} & \multicolumn{1}{l}{Final tune} &  \\
\midrule
\PARP{64}  & 1.0 & 1.3 & ISR scale factor for $\alpha_S$ \\
\PARP{71}  & 4.0 & 2.0 & max. virtuality (non-s-channel) \\
\PARP{78}  & 0.03 & 0.17 & colour reconnection in FSR \\
\PARP{79}  & 2.0 & 1.18 & beam remnant x enhancement \\
\PARP{80}  & 0.1 & 0.01 & beam remnant breakup suppression \\
\PARP{82}  & 2.0 & 1.85 & $\pT^0$ at reference $E_\text{cm}$ \\
\PARP{83}  & 1.8 & 1.8 & matter distribution \\
\PARP{90}  & 0.16 & 0.22 & $\pT^0$ energy evolution \\
\PARP{91}  & 2.0 & 2.0 & intrinsic $k_\perp$ \\
\PARP{93}  & 5.0 & 7.0 & intrinsic $k_\perp$ cut-off \\
\bottomrule
\end{tabular}

\bigskip
\begin{tabular}{lrl}
\toprule
Switch  & Value & Effect \\
\midrule
\MSTJ{41}  & 12 & switch on $\pT$-ordered shower \\
\MSTP{51}  &  7 & use CTEQ5L \\
\MSTP{52}  &  1 & use internal PDF set \\
\MSTP{70}  &  2 & model for smooth $\pT^0$ \\
\MSTP{72}  &  0 & FSR model \\
\MSTP{81}  & 21 & turn on multiple interactions (new model) \\
\MSTP{82}  &  5 & model of hadronic matter overlap \\
\MSTP{88}  &  0 & quark junctions $\to$ diquark/Baryon model \\
\MSTP{95}  &  6 & colour reconnection \\
\bottomrule
\end{tabular}
\end{center}
\caption{Tuned parameters (upper table) and switches (lower table) for the
         underlying event using the $\pT$-ordered shower.}
\label{tab:params-uept}
\end{table}

\section{Conclusions}

The \rivet{} and \professor{} tools are in a state where they can be
used for real tunings and the tuning of \pythiasix{} has been a
significant success. At and around the Perugia workshop a bunch of new
tunes appeared on the market: Our \professor{} tunes, Peter Skand's
Perugia tunes (which are based on our flavour and fragmentation
parameters), and combinations of the well established Rick Field tunes
with our new flavour and fragmentation settings which even improve the
agreement with data at the \tevatron{}. All these tunes are directly
available through the PYTUNE routine in \pythia{}~6.420 or later.

We strongly encourage the \lhc{} experiments to use one of these tunings
instead of spending their valuable time on trying to tune themselves.
Monte Carlo tuning requires a sound understanding of the models and of
the data, and a very close collaboration with the generator authors. In
the current situation we highly recommend the use of either Peter
Skands' Perugia tune or our new tune if the user wants to go for the new
MPI model, or a tune like DWpro or our tune of the virtuality-ordered
shower for a more conservative user who wants to use a well-proven
model.

\section*{Acknowledgements}

We want to thank the organisers of the MPI@LHC workshop for this very
productive and enjoyable meeting, and we are looking forward to its
continuation. Our work was supported in part by the MCnet European Union
Marie Curie Research Training Network, which provided funding for
collaboration meetings and attendance at research workshops such as
MPI@LHC'08. Andy Buckley has been principally supported by a Special
Project Grant from the UK Science \& Technology Funding Council. Hendrik
Hoeth acknowledges a MCnet postdoctoral fellowship.

\begin{footnotesize}
\bibliographystyle{mpi08}
{\raggedright
\bibliography{mpi08}
}
\end{footnotesize}
\end{document}